\documentclass[jcp,aps,showpacs,twocolumn,supergroupedaddress,epsfig,amsmath,amssymb,eqsecnum]{revtex4}
\usepackage{epsfig,amsmath,amsthm,amssymb,bm,epsf,graphics,psfrag,verbatim,subfigure,framed}
\usepackage[makeroom]{cancel}
\usepackage{mathtools}
\usepackage[usenames,dvipsnames]{color}
\usepackage{empheq}
\usepackage{bbm}
\usepackage{mathrsfs}
\usepackage{amsfonts}
\usepackage{color} 
\usepackage{pbox}
\usepackage{xcolor} 
\usepackage{multirow}
\usepackage{hyperref}
\def\tsigma{\widetilde{\sigma}}

\def\nv{\mathbf{n}}

\newcommand{\tk}{{\widetilde k}}

\newcommand{\tG}{{\widetilde G}}
\newcommand{\tomega}{{\widetilde \omega}}

\newcommand{\be}{\begin{equation}}
\newcommand{\ee}{\end{equation}}
\newcommand{\ba}{\begin{eqnarray}}
\newcommand{\ea}{\end{eqnarray}}
\newcommand{\bw}{\begin{widetext}}
\newcommand{\ew}{\end{widetext}}

\newcommand{\rv}{{\mathbf{r}}}

\newcommand{\kv}{{\bf k}}
\newcommand{\pv}{{\bf p}}

\newcommand{\Dv}{{\bf D}}

\newcommand{\bing}[1]{\textcolor{black}{#1}}

\begin{document}

\title{An excited atom interacting with a Chern insulator: \\ 
towards a far-field resonant Casimir-Polder repulsion}
\author{Bing-Sui Lu$^1$}
\email{BingSui.Lu@xjtlu.edu.cn}
\author{Khatee Zathul Arifa$^2$}
\author{Martial Ducloy$^3$}
\affiliation{$^1$Department of Physics, Xi'an Jiaotong-Liverpool University, Suzhou, Jiangsu 215123, China
\\
$^2$Department of Physics, University of Wisconsin-Madison, Madison, WI 53706-1390, USA\\
$^3$Laboratoire de Physique des Lasers, Universit\'{e} Sorbonne Paris-Nord, 93430 Villetaneuse, France}
\date{\today}

\begin{abstract}
We investigate the resonant Casimir-Polder interaction of an excited atom which has a single (electric dipole) transition with a Chern insulator, using the approach of quantum linear response theory. 
The Chern insulator has a nonzero, time reversal symmetry breaking Hall conductance, leading to an additional contribution to the resonant Casimir-Polder interaction which depends on the coupling between the Hall conductance and the circular polarization state of the atomic transition. 
We find that the resonant Casimir-Polder shift can be significantly enhanced if the atomic de-excitation frequency is near a value associated with a van Hove singularity of the Chern insulator. Furthermore, we find that the resonant Casimir-Polder force can become monotonically decaying and repulsive for a relatively large atom-surface distance. This happens if the atomic dipole transition is right (left) circularly polarized and the Chern number of the Chern insulator is $-1$ ($+1$), and the atomic de-excitation energy is comparable to the bandgap energy of the Chern insulator. This has potential implications for the design of atom-surface interaction which can be tuned repulsive over a relatively large range of separations.   
\end{abstract}

\maketitle

\section{Introduction} 

An atom can interact with a surface via the so-called Casimir-Polder (CP) interaction~\cite{CP-paper,ducloy-review,laliotis2021}. 
In the near-field (or nonretarded) region, which is defined to be smaller than a characteristic distance set by the ratio of the speed of light to the resonant transition frequency of the atom, the CP interaction becomes the van der Waals interaction, which can be physically regarded as the dispersive electrostatic interaction of a dipole with its image dipole across the dielectric surface. The near-field interaction potential decays 
as $-hC_3(|i\rangle)/z_0^3$, where $C_3(|i\rangle)$ is the so-called van der Waals coefficient for an atom in the $i$th atomic state, $h$ is the Planck constant, and $z_0$ is the distance between the atom and the surface. 
The CP interaction between an atom and a surface has been the subject of numerous studies, because of its relevance to technological applications~\cite{bordag}, experiments on atomic gases~\cite{shimizu2001,oberst2005,harber2005}, and the search for constraints on non-Newtonian forces~\cite{wolf2007,chwedenczuk2010,bennett2019}. In recent years, owing to the surge of interest in exotic materials, the CP interaction has also been studied in the context of metamaterials~\cite{chan2018}, graphene~\cite{ribeiro2013,churkin2010,cysne2014,khusnutdinov2016}, Chern-Simons surfaces~\cite{marachevsky2010,buhmann2018}, axionic media~\cite{crosse2015,fuchs2017}, and topological photonics~\cite{silveirinha2018}. 

Although in many instances the CP interaction is attractive (for example, a ground state atom interacting with a homogeneous planar surface which is reciprocal, the whole system being maintained at thermal equilibrium), there are situations where the interaction can become repulsive, for example, systems with a high degree of anisotropy~\cite{eberlein2011,shajesh2012,milton2012} and thermally out-of-equilibrium systems~\cite{antezza2005}.  
A repulsive CP force can also be realized for an atom in the excited state, for which the CP potential has a resonant contribution that is enhanced at frequencies near the characteristic frequency of the surface polariton. The CP force oscillates with distance from the surface, thus the force can become repulsive at certain distances~\cite{chance1975,wylie-sipe2,fichet1995,failache1999,failache2003}. 
For a birefringent surface near an excited atom which is in resonance with the surface polariton mode, the repulsive CP force can also be tuned by rotating the principal optic axis of the birefringent substrate~\cite{gorza2001}. Finally, it was found~\cite{crosse2015,fuchs2017} for an axionic topological medium with a \emph{nondispersive} Hall conductance that the CP force can tuned from attraction to repulsion, by changing the sign of the axionic coupling. 


In this paper, we investigate the Casimir-Polder interaction between an atom in an excited state and a Chern insulator~\cite{qwz2006,asboth2016,lu2021}, accounting for the full frequency dispersion of the conductivity response of the Chern insulator. 
Such an investigation is motivated by the existence of three broad types of dispersive behavior that the conductivity can exhibit depending on the frequency regime of the photon, which leads to qualitative differences in the spontaneous emission behavior of a neighboring excited atom. We study the resonant Casimir-Polder interaction corresponding to each of the frequency regimes, and find that the resonant Casimir-Polder force can be greatly enhanced if the excited atom is resonant at frequencies associated with a van Hove singularity of the Chern insulator. We consider the Chern insulator because it is one of the simpler examples of a nonreciprocal Hall surface, which enables one to investigate how the presence of the nonreciprocity makes possible the emergence of an effectively monotonically decaying and hence longer-ranged repulsive resonant Casimir-Polder force. Such a phenomenon is predicted for certain sign combinations of the Chern number and the angular momentum change associated with the atomic dipole transition, and does not occur if the surface is made of a reciprocal material. 

Our methodology is based on the linear response approach pioneered by Refs.~\cite{wylie-sipe1} and \cite{wylie-sipe2}, which has seen applications to systems consisting of an atom above a birefringent medium~\cite{gorza2001}, an atom above a dielectric surface at finite temperature~\cite{gorza2006}, an atom inside an optical cavity~\cite{nha1996}, and spontaneous emission from an atom above a Chern insulator~\cite{lu2020}. Our study can be regarded as a sequel to Ref.~\cite{lu2020} and follows esssentially the same assumptions adopted in that work, {\it i.e.}, we consider an atom with a single (electric dipole) transition and assume atom-surface separations to be large enough such that the wavefunction of the atom has no appreciable overlap with that of the surface. We also assume that the temperature of the system is at or close to zero Kelvin, so that thermal effects can be neglected. 
To capture the full dispersion of the conductivity response, we use Kubo's linear response theory~\cite{czycholl2017}.

\section{Theoretical method} 

\subsection{Model of a Chern insulator}

A Chern insulator is a two-dimensional system exhibiting the quantum anomalous Hall effect (QAHE)~\cite{cayssol2013,bernevig2013,weng2015,liu2016,ren2016,zhang2016}. The QAHE is well known for giving rise to a Hall conductance whose static limit is integer quantised to $C e^2/h$, where $e = 1.602\times 10^{-19} \, {{\rm C}}$ is the electron charge, $h = 6.626\times 10^{-34} \, {{\rm J.s}}$ is the Planck constant, and $C$ is an integer called the Chern number. Such a system is also characterized by the vanishing of the static limit of the longitudinal conductance. 
A simple model which can account for these features as well as the effect of dispersion is the Qi-Wu-Zhang (QWZ) model~\cite{qwz2006}, described by 
\be
\label{H2band}
H = {\bf{d}}(\kv) \cdot {\bm{\sigma}} = d_x(\kv) \sigma_x + d_y(\kv) \sigma_y + d_z(\kv) \sigma_z. 
\ee
Here, $\sigma_x$, $\sigma_y$ and $\sigma_z$ are the Pauli matrices, $d_x(\kv) = t \sin k_x a$, $d_y(\kv) = t \sin k_y a$, and $d_z(\kv) = t (\cos k_x a + \cos k_y a) + u$, where $t$ denotes the hopping amplitude and $u$ is the band gap opening induced by the magnetization on the Chern insulator surface. It is known that this model generates a Chern number of $1$ ($-1$) for $0<u<2t$ ($-2t < u < 0$). In the rest of our paper, we consider the case $|u| = t$, which gives rise to a particularly pronounced van Hove singularity~\cite{lu2020}. 
From the QWZ model, the conductivity tensor can be calculated using the Kubo formula. The presence of a nonzero $u$ breaks time-reversal symmetry, which leads to a nonzero Berry curvature and thus a nonzero Hall conductance. The Hall conductance is represented by the terms $\sigma_{xy}$ and $\sigma_{yx}$
in the conductivity tensor, and importantly the reciprocity is broken as $\sigma_{xy} = - \sigma_{yx}$~\cite{lu2020}. 

\subsection{Shifts of the atomic energy levels}

The presence of the Chern insulator results in shifts to the energy levels of an atom placed in its vicinity, and the energy level shift also depends on the separation distance between the atom and the Chern insulator. Consequently, the atom feels a force. 
The shift in the energy level can be derived from second-order perturbation theory~\cite{wylie-sipe1,wylie-sipe2}; such a calculation involves the use of the dyadic electromagnetic Green function which relates the change in the expectation value of the displacement field at time $t$ to the dipole source at time $t'$ which gives rise to it~\cite{wylie-sipe1}. 
The Green function in the dielectric half-plane which contains the atom can be decomposed into the sum of two contributions, {\it i.e.}, a bulk contribution, which we denote by $\mathbb{G}^{0}$ and describes an atom in free space, and a contribution by reflected waves, which we denote by $\mathbb{G}^{R}$ and represents the modification coming from waves reflecting off the Chern insulator. 
\begin{figure}[h]
\centering
  \includegraphics[width=0.43\textwidth]{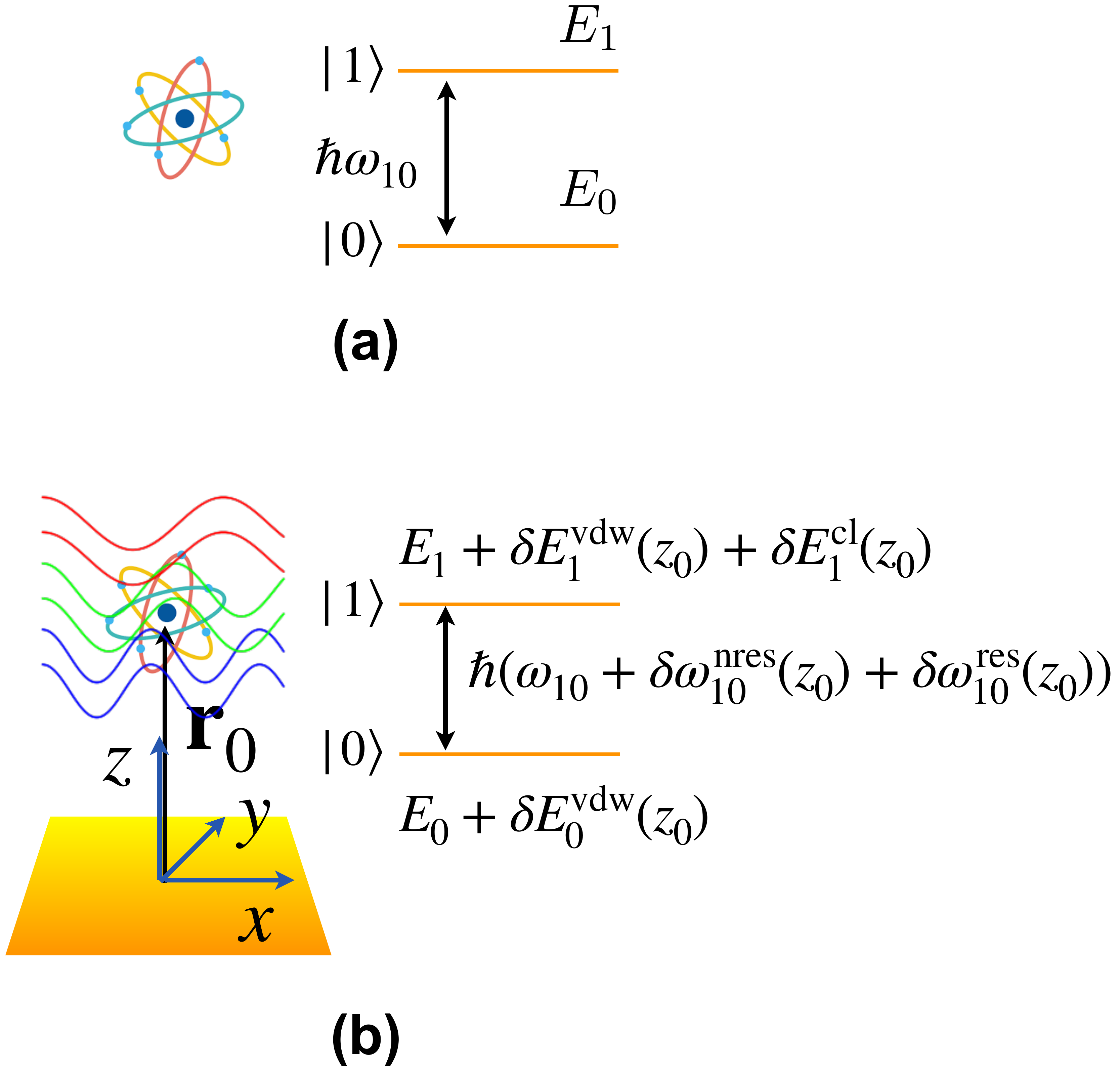}
  \caption{(a)~An isolated two-level atom in the absence of radiation. There are two levels, a ground state $|0\rangle$ with energy $E_0$ and an excited state $|1\rangle$ with energy $E_1$. (b)~A two-level atom in the presence of radiation (wavy lines) and positioned near a two-dimensional surface (colored gold). The energy of $|0\rangle$ ($|1\rangle$) is shifted to $E_0 + \delta E_{0}^{\rm{vdw}}$ ($E_1 + \delta E_{1}^{\rm{vdw}} + \delta E_1^{\rm{cl}}$). For the coordinate frame, we choose the $x$ and $y$ axes ($z$ axis) to be parallel (parallel) to the surface; $\rv_0 = (0,0,z_0)^{{\rm T}}$ (where the superscript ${{\rm T}}$ denotes the transpose) is the position vector of the atom. The quantities $\delta E_{0/1}^{\rm{vdw}}$, $\delta E_1^{\rm{cl}}$, $\omega_{10}^{{\rm nres}}$ and $\omega_{10}^{{\rm res}}$ are defined in the main text and given by Eqs.~(\ref{marimo}) and (\ref{deltaomega}).} 
  \label{fig:geometry}
\end{figure}

For our purpose, we consider an atom with a single (electric dipole) transition which connects the ground state (with $J=0$) with a higher energy state (with $J=1$ and substates with $m_J = -1, 0, 1$). We denote the ground state by $|0\rangle$ and the excited state by $|1\rangle$ (cf. Fig.~\ref{fig:geometry}). 
The energy of the atomic ground state can be shifted both by electromagnetic vacuum fluctuations of the free space (and leads to the Lamb shift) and polarization fluctuations of the neighboring material surface; following the convention of Ref.~\cite{wylie-sipe1}, we denote the surface-induced energy shift of the ground state by $\delta E_0^{{\rm vdw}}$. There is a similar surface-induced shift in the energy of the excited state, and this shift has two contributions: 
\begin{itemize}
\setlength{\parskip}{0pt} \setlength{\itemsep}{0pt plus 1pt} 
\item the energy of polarization of the atom's electron cloud by electromagnetic vacuum fluctuations, which we denote by $\delta E_1^{{\rm vdw}}$; and 
\item a contribution that appears only for an atom in the excited state and which in the nonretarded regime can be interpreted as originating from the interaction of a dipole with its image dipole; we denote this contribution by $\delta E_1^{{\rm cl}}$. 
\end{itemize} 
We call the surface-induced shift the \emph{Casimir-Polder shift.}  
Using perturbation theory, we find (App.~\ref{sec:app1}) that the Casimir-Polder shift for the state $|m\rangle$ 
is given by 
$\delta E_m' = \delta E_m^{{\rm vdw}} + \delta E_m^{{\rm cl}}$, where 
\ba
\delta E_1^{{\rm vdw}} 
&\!\!\!\!\!\!=\!\!\!\!&
\frac{1}{\pi} 
\int_0^\infty \!\!\!\!\! d\xi 
\Big(
\frac{\omega_{10} \big( G_{ab}(\rv_0,\rv_0; i\xi) + G_{ba}^\ast(\rv_0,\rv_0; i\xi) \big)}{2(\omega_{10}^2 + \xi^2)}
\nonumber\\
&&\,\,\,\,-
\frac{\xi \big( G_{ab}(\rv_0,\rv_0; i\xi) - G_{ba}^\ast(\rv_0,\rv_0; i\xi) \big)}{2i(\omega_{10}^2 + \xi^2)}
\Big)
\mu_a^{10} \mu_b^{01},
\nonumber\\
\delta E_0^{{\rm vdw}} 
&\!\!\!\!\!\!=\!\!\!\!&
\frac{1}{\pi} 
\int_0^\infty \!\!\!\!\! d\xi 
\Big(
\frac{\omega_{01} \big( G_{ab}(\rv_0,\rv_0; i\xi) + G_{ba}^\ast(\rv_0,\rv_0; i\xi) \big)}{2(\omega_{01}^2 + \xi^2)}
\nonumber\\
&&\,\,\,\,-
\frac{\xi \big( G_{ab}(\rv_0,\rv_0; i\xi) - G_{ba}^\ast(\rv_0,\rv_0; i\xi) \big)}{2i(\omega_{01}^2 + \xi^2)}
\Big)
\mu_a^{01} \mu_b^{10},
\nonumber\\
\delta E_1^{{\rm cl}} 
&\!\!\!\!=\!\!\!\!&
- 
\frac{1}{2} \mu_a^{10} \mu_b^{01} 
\big( G_{ab}(\rv_0,\rv_0; \omega_{10}) 
+ G_{ba}^\ast(\rv_0,\rv_0; \omega_{10}) \big), 
\nonumber\\
\delta E_0^{{\rm cl}} 
&\!\!\!\!=\!\!\!\!& 
0.
\label{marimo}
\ea
The quantity $\omega_{10} \equiv \omega_1 - \omega_0 \equiv  (E_1 - E_0)/\hbar$ is the frequency of transition between the unperturbed excited state $|1\rangle$ and the ground state $|0\rangle$. 
Following the terminology in the literature, we call $\delta E_1^{{\rm cl}}$ the \emph{resonant} Casimir-Polder shift, and call $\delta E_0^{{\rm vdw}}$ and $\delta E_1^{{\rm vdw}}$ \emph{nonresonant} Casimir-Polder shifts. 
The symbol ${\bm{\mu}}^{10} \equiv \langle 1 | {\bm{\mu}} | 0 \rangle$ denotes the atomic dipole transition matrix element from $| 0 \rangle$ to $| 1 \rangle$, and ${\bm{\mu}} = -e\, \hat{q}\, \nv$ denotes the electric dipole operator of the atom, where $\nv$ specifies the polarization direction of a photon undergoing an electric dipole transition, and $\hat{q}$ is the position operator. In Sec.~\ref{sec:CP-shifts}, we shall consider a few different examples of the photon polarization, \emph{i.e.}, perpendicular to the Chern insulator surface, parallel to the surface, and the two orthogonal circular polarizations in the plane of the Chern insulator.  
Adopting an oscillator model for the atom~\cite{wylie-sipe2}, the atomic dipole transition matrix element becomes 
$\mu_a^{10} = \mu_a^{01*}
= \mu \, n_a$, 
where $\mu \equiv -e ( \hbar/(2m\omega_{10}) )^{1/2}$. 
As we are interested in how the presence of the Chern insulator modifies the atom's energy level shifts, we focus on the reflected wave contribution to the dyadic Green's function, {\it i.e.}, $\mathbb{G}^R$. 

\bing{The Casimir-Polder shift also changes the atomic transition frequency from the unperturbed value of $\omega_{10}$ to $\omega_{10} + \delta\omega_{10}$, where $\delta\omega_{10} \equiv (\delta E_1^{{\rm cl}} + \delta E_1^{{\rm vdw}} - \delta E_0^{{\rm vdw}})/\hbar$ (see Fig.~\ref{fig:geometry}). We can express $\delta\omega_{10}$ as the sum of a resonant  shift $\delta\omega_{10}^{{\rm res}}$ and a nonresonant shift $\delta\omega_{10}^{{\rm nres}}$: 
\ba
\label{deltaomega}
\delta\omega_{10} 
&\!\!=\!\!& 
\delta\omega_{10}^{{\rm res}} + \delta\omega_{10}^{{\rm nres}}, 
\\
\delta\omega_{10}^{{\rm res}} &\equiv& \frac{\delta E_1^{{\rm cl}}}{\hbar}, \,\, 
\delta\omega_{10}^{{\rm nres}} \equiv \frac{1}{\hbar} \left( \delta E_1^{{\rm vdw}} - \delta E_0^{{\rm vdw}} \right)
\nonumber
\ea
As the transition rate $R_{10}^{(0)}$ of an atom in free space has dimensions of frequency, where~\cite{wylie-sipe1} 
\be
\label{R10}
R_{10}^{(0)} = \frac{4\omega_{10}^3 \mu^2}{3 \hbar c^3}, 
\ee
and $\delta E_0^{{\rm cl}} = 0$, it is also convenient to consider the behavior of the dimensionless frequency shifts: 
\begin{subequations}
\label{dimfreqshifts}
\ba
\delta\widetilde{\omega}_{10} 
&\!\!=\!\!& 
\delta\widetilde{\omega}_{10}^{{\rm res}} + \delta\widetilde{\omega}_{10}^{{\rm nres}}, 
\\
\delta\widetilde{\omega}_{10}^{{\rm res}} 
&\equiv& 
\frac{\delta\omega_{10}^{{\rm res}}}{R_{10}^{(0)}}, 
\,\, 
\delta\widetilde{\omega}_{10}^{{\rm nres}} 
\equiv 
\frac{\delta\omega_{10}^{{\rm nres}}}{R_{10}^{(0)}}.  
\ea
\end{subequations}
Using Eqs.~(\ref{marimo}) and (\ref{R10}) we can express 
\be
\label{formular}
\delta\widetilde{\omega}_{10}^{{\rm res}} 
= - \frac{3}{8} n_a n_b^\ast 
\left( 
\tG_{ab}(\omega_{10}) + \tG_{ba}^\ast(\omega_{10})
\right). 
\ee
In Sec.~\ref{sec:CP-shifts}, we study the resonant contribution to the Casimir-Polder shift in the atomic transition frequency, $\delta\tomega_{10}^{{\rm res}}$, for the following polarizations of a photon undergoing an electric dipole transition from the excited state: 
\begin{itemize}
\setlength{\parskip}{0pt} \setlength{\itemsep}{0pt plus 1pt} 
\item right circular photon polarization with the quantization axis perpendicular to the surface of a Chern insulator with $C = 1$; and 
\item right circular photon polarization with the quantization axis perpendicular to the surface of a Chern insulator with $C = -1$. 
\item photon polarization perpendicular to the surface of a Chern insulator with $C=1$;  
\item photon polarization parallel with the surface of a Chern insulator with $C = 1$; 
\end{itemize} 
As a circular photon polarization breaks time-reversal symmetry, it can couple to the Hall conductance of the Chern insulator. This coupling gives rise to an additional channel in the resonant Casimir-Polder shift which breaks reciprocity. As we shall see, depending on the relative orientation of the circular polarization and the Hall current, the resonant Casimir-Polder shift can exhibit quite different behaviors.  }

\subsection{Electromagnetic Green tensor}
\label{sec:green-tensor}

As we are interested in studying the behavior of an excited atom with a dipole transition polarized either perpendicular to or parallel with the plane of the Chern insulator, only the xx, yy, xy, yx and zz components of the reflection Green tensor will contribute to the Casimir-Polder shift (here, we take the surface of the Chern insulator to be the xy plane and the z direction to be parallel to the surface normal; see Fig.~\ref{fig:geometry}). 
For a two-dimensional planar surface with longitudinal conductivity $\sigma_{xx} = \sigma_{yy}$ and Hall conductivity $\sigma_{xy} = -\sigma_{yx}$, these components are~\cite{lu2020}
\ba
&&G^R_{xx}(\rv_0, \rv_0, \omega) 
= 
G^R_{yy}(\rv_0, \rv_0, \omega) 
\nonumber\\
&=&
\frac{i}{2} \Big( \frac{\omega}{c} \Big)^2 \! 
\int_{0}^\infty \!\!\!\! dk_\parallel \frac{k_\parallel}{k_z} \, e^{2 i k_z z} 
\big( 
r_{ss} - \frac{c^2 k_z^2}{\omega^2} r_{pp} 
\big), 
\nonumber\\
&&G^R_{xy}(\rv_0, \rv_0, \omega) 
= 
- G^R_{yx}(\rv_0, \rv_0, \omega) 
\nonumber\\
&=&
\frac{i}{2} \Big( \frac{\omega}{c} \Big) \! 
\int_{0}^\infty \!\!\!\! dk_\parallel k_\parallel e^{2 i k_z z} 
\left( r_{ps} + r_{sp} \right), 
\nonumber\\
&&G^R_{zz}(\rv_0, \rv_0, \omega) 
=
i \! 
\int_{0}^\infty \!\!\!\! dk_\parallel \frac{k_\parallel^3}{k_z} \, e^{2 i k_z z}  
r_{pp}.
\label{G-disp}
\ea
In our expressions for the Green tensor, we have taken the source and field points to be coincident, {\it i.e.}, $\rv= \rv_0$, as this is the configuration that we are interested in. 
We define $\rv_\parallel = (x, y)^{{\rm T}}$ (where the superscript ${{\rm T}}$ denotes the transpose), 
$\kv_\parallel = (k_x, k_y)^{{\rm T}}$, 
and $k_z = ((\omega/c)^2 - k_\parallel^2)^{1/2}$. 
The Green tensor in Eq.~(\ref{G-disp}) involves a factor of $1/k_z$, which may make the integral appear to be singular. This factor comes from the Weyl identity, which requires ${{\rm Re}} \, [k_z] > 0$ and ${{\rm Im}} \, [k_z] > 0$ for its fulfilment~\cite{radiation-condition}; thus, the integral is actually \emph{not} singular.
If we define the dimensionless parameters $\widetilde{k}_z \equiv ck_z/\omega_{10}$,  $\tk_\parallel \equiv ck_\parallel/\omega_{10}$ and $\eta \equiv 2 \omega_{10} z_0 / c$, we can obtain a dimensionless Green tensor
$\widetilde{G}_{\mu\nu}^R \equiv (c/\omega_{10})^3 G_{\mu\nu}^R$, where 
\begin{subequations}
\label{dyadic}
\ba
\label{Gxx}
&&\widetilde{G}_{xx}^R(\rv_0,\rv_0;\omega) = \widetilde{G}_{yy}^R(\rv_0,\rv_0;\omega) 
\\
&\!\!=\!\!&
\frac{i}{2} 
\int_0^\infty \!\!\! d\tk_\parallel (\tk_\parallel/\tk_z)
\big( (\omega/\omega_{10})^2 
 r_{ss} - \tk_z^2 r_{pp} 
\big)
e^{i \tk_z \eta},
\nonumber
\\
&&\widetilde{G}_{xy}^R(\rv_0,\rv_0;\omega) = - \widetilde{G}_{yx}^R(\rv_0,\rv_0;\omega) 
\label{Gxy}
\\
&\!\!=\!\!& 
\frac{i}{2} 
\int_0^\infty \!\!\! d\tk_\parallel \tk_\parallel (\omega/\omega_{10}) (r_{ps} + r_{sp}) e^{i \tk_z \eta},
\nonumber\\
&&\widetilde{G}_{zz}^R(\rv_0,\rv_0;\omega) = 
i  
\int_0^\infty \!\!\! d\tk_\parallel (\tk_\parallel^3/\tk_z) r_{pp} e^{i \tk_z \eta}. 
\label{Gzz}
\ea
\end{subequations}
Here, $r_{ss}, r_{ps}, r_{sp}$ and $r_{pp}$ are the reflection coefficients respectively for an incident s-polarized wave getting reflected as an s-polarized wave, an incident p-polarized wave getting reflected as an s-polarized wave, an incident s-polarized wave getting reflected as a p-polarized wave, and an incident p-polarized wave getting reflected as a p-polarized wave.
These coefficients are given by~(cf. App.~\ref{app:fresnel})~\cite{footnote1}
\ba
\label{r-coeffs}
r_{pp} &\!\!=\!\!& 
\frac{1}{\Delta} \big( \widetilde{\sigma}_{xx}^2 + \widetilde{\sigma}_{xy}^2 + (\omega_{10}/\omega) \widetilde{k}_z \widetilde{\sigma}_{xx} \big), 
\\
r_{ss} &\!\!=\!\!& 
-\frac{1}{\Delta} \big( \widetilde{\sigma}_{xx}^2 + \widetilde{\sigma}_{xy}^2 + (\omega/\omega_{10}) \widetilde{k}_z^{-1} \widetilde{\sigma}_{xx} \big), 
\nonumber\\
r_{ps} &\!\!=\!\!& r_{sp} = - \frac{\widetilde{\sigma}_{xy}}{\Delta}, 
\nonumber\\
\Delta &\!\!\equiv\!\!& 1 + \big( (\omega_{10}/\omega) \widetilde{k}_z + (\omega/\omega_{10}) {\widetilde{k}_z^{-1}} \big) \widetilde{\sigma}_{xx} + \widetilde{\sigma}_{xx}^2 + \widetilde{\sigma}_{xy}^2.
\nonumber
\ea
In the above, $\widetilde{\sigma}_{\mu\nu} \equiv 2\pi\sigma_{\mu\nu}/c$, and $\sigma_{\mu\nu}$ is the conductivity tensor. The conductivity tensor is derived in Ref.~\cite{lu2020} using the Kubo formula, and we refer the reader to that paper for the detailed expressions. 
When an excited atom de-excites, the emitted photon can get absorbed by the Chern insulator, and as discussed in the same reference, depending on the frequency $\omega_{10}$ of the absorbed photon, the conductivity response can belong to one of the following three regimes: 
\begin{itemize}
\setlength{\parskip}{0pt} \setlength{\itemsep}{0pt plus 1pt} 
\item a \emph{low frequency regime}, for which $\hbar\omega_{10} < 2(2t-|u|)$, {\it i.e.}, the energy of the absorbed photon is smaller than the minimum value of the insulator band gap; 
\item an \emph{intermediate frequency regime}, for which $2(2t-|u|) \leq \hbar\omega_{10} \leq 2(2t+|u|)$, {\it i.e.}, the energy of the absorbed photon is within the insulator band gap; 
\item a \emph{high frequency regime}, for which $\hbar\omega_{10} > 2(2t+|u|)$, {\it i.e.}, the energy of the absorbed photon is greater than the largest insulator band gap value.
\end{itemize} 
For the choice $|u| = t$, the low frequency regime corresponds to $\hbar\omega_{10} < 2t$, the intermediate frequency regime corresponds to $2t \leq \hbar\omega_{10} \leq 6t$, and the high frequency regime corresponds to $\hbar\omega_{10} > 6t$. 
These different regimes are characterized by qualitatively distinct behaviors of the conductivity response, and give rise, for example, to correspondingly distinct asymptotics in the spontaneous emission behavior of the atom. For example, in the low and high frequency regimes, the real part of the longitudinal conductance and the imaginary part of the Hall conductance vanish~\cite{lu2020}. The surface-induced correction to the transition rate was found to diverge in the near-field limit if $\omega_{10}$ lies within the intermediate frequency regime, whereas the surface-induced correction is finite if $\omega_{10}$ lies within the low and high frequency regimes. The longitudinal conductance is nonzero in the intermediate frequency regime owing to the promotion of valence band electrons into the conduction band, whereas it is zero for the other two frequency regimes. On the other hand, the Hall conductance tends to a nonzero (quantized) value in the zero frequency limit of the low frequency regime, whilst the same conductance vanishes in the infinite frequency limit of the high frequency regime. 


\section{Results and Discussion}
\label{sec:CP-shifts}

\subsection{circularly polarized dipole transition}

\begin{figure}[b]
  \centering
\includegraphics[width=0.45\textwidth]{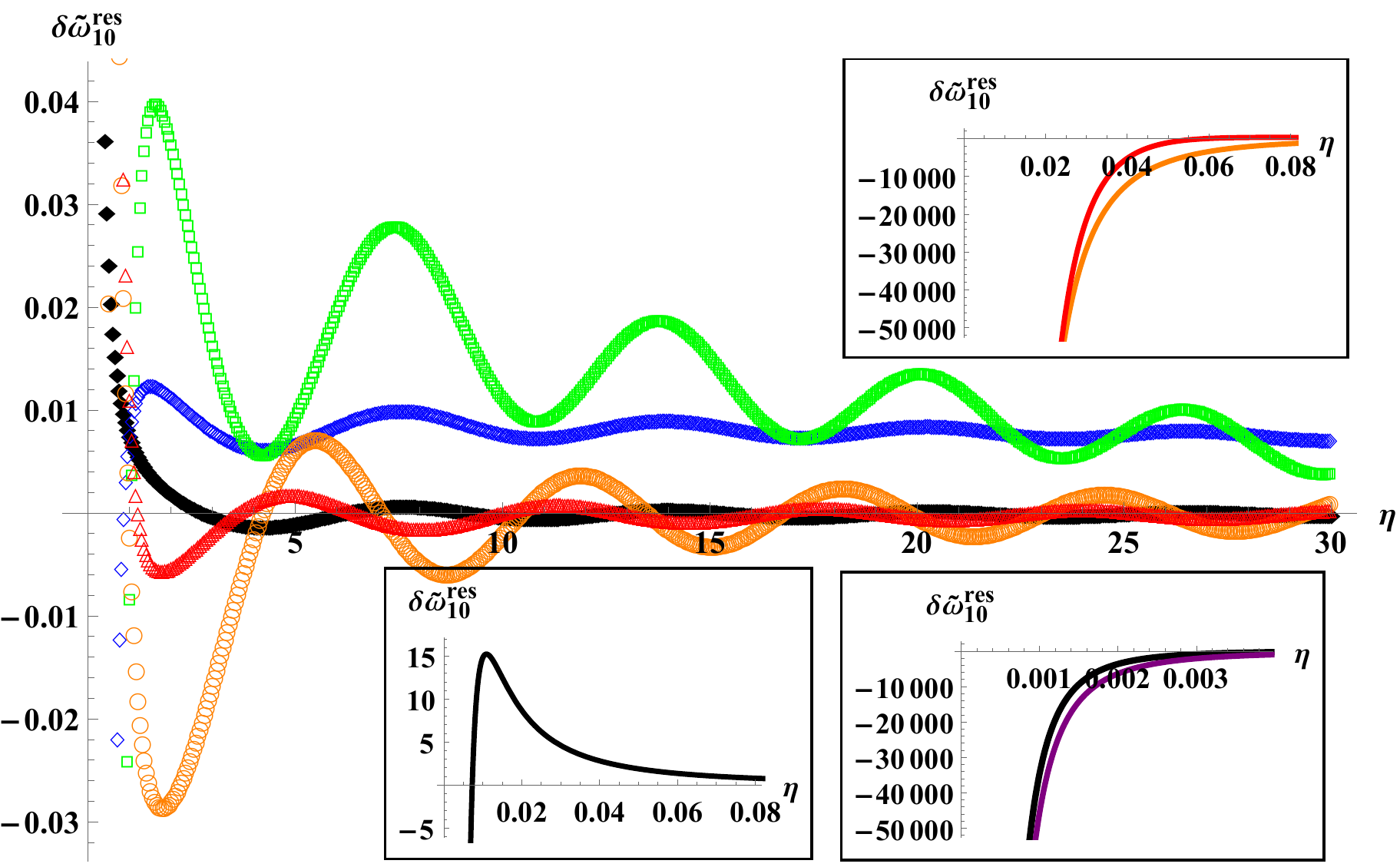}
\caption{Resonant Casimir-Polder shift $\delta\widetilde{\omega}_{10}^{{\rm res}}$ (vertical axis) as a function of dimensionless atom-surface separation distance $\eta \equiv 2(\omega_{10}/c)z_0$ (horizontal axis) for an atomic dipole transition with a right circular polarization parallel with the surface of a $C = 1$ Chern insulator, and $u/t = 1$. 
Legend: ~nondispersive (black, filled diamond), $\hbar \omega_{10}/t = 1$ (blue, diamond), $\hbar \omega_{10}/t = 1.9$ (green, square), $\hbar \omega_{10}/t = 2.1$ (orange, circle), and $\hbar \omega_{10}/t = 3$ (red, triangle). 
Top inset: Behavior of $\delta\widetilde{\omega}_{10}^{{\rm res}}$ in the near-field regime ($\eta \ll 1$) for the cases $\hbar \omega_{10}/t = 2.1$ and 3. 
Bottom left inset: Behavior of $\delta\widetilde{\omega}_{10}^{{\rm res}}$ in the near-field regime ($\eta \ll 1$) for the case of nondispersive conductivity. 
Bottom right inset: Behavior of nondispersive $\delta\widetilde{\omega}_{10}^{{\rm res}}$ in the near-field regime ($\eta \ll 1$) for $C=1$ (black) and $C=-1$ (purple) Chern insulators.}
   \label{resonantCPcirc1}
  \end{figure}

\begin{figure}[h]
  \centering
    \includegraphics[width=0.45\textwidth]{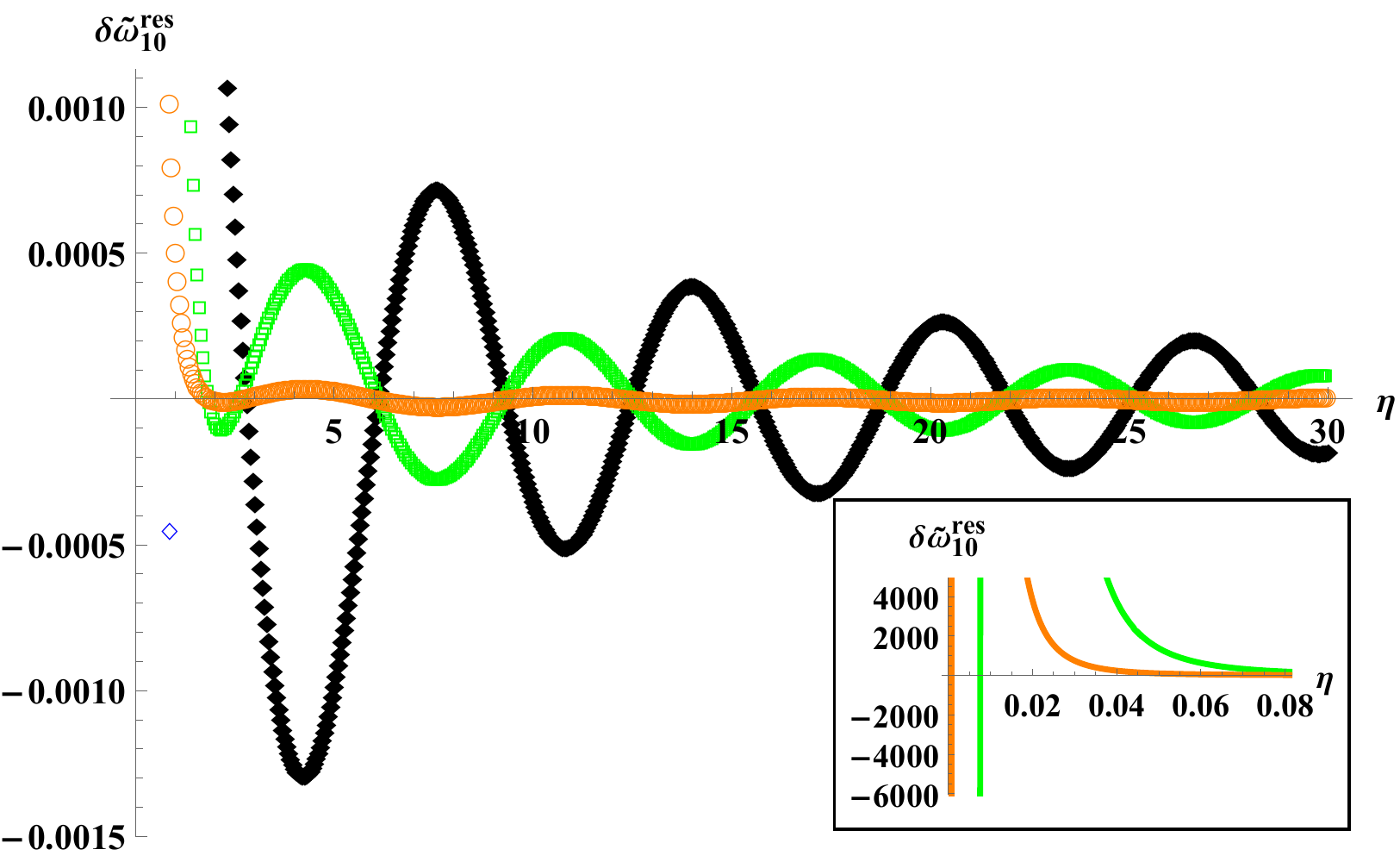}
    \caption{Resonant Casimir-Polder shift $\delta\widetilde{\omega}_{10}^{{\rm res}}$ (vertical axis) as a function of dimensionless atom-surface separation distance $\eta \equiv 2(\omega_{10}/c)z_0$ (horizontal axis) for an atomic dipole transition with a right circular polarization aligned parallel with the surface of a $C = 1$ Chern insulator, and $u/t = 1$. We show behaviors both corresponding to the case of nondispersive conductivity (black, filled diamond) as well as dispersive conductivity with $\hbar\omega_{10}/t = 10$ (green, square) and $\hbar\omega_{10}/t = 100$ (orange, circle). Top inset: Behavior of $\delta\widetilde{\omega}_{10}^{{\rm res}}$ in the near-field regime ($\eta \ll 1$) for $\hbar \omega_{10}/t = 10$ (green, square) and $100$ (orange, circle).}
   \label{resonantCPcirc2}
\end{figure}

\begin{figure}[h]
  \centering
    \includegraphics[width=0.45\textwidth]{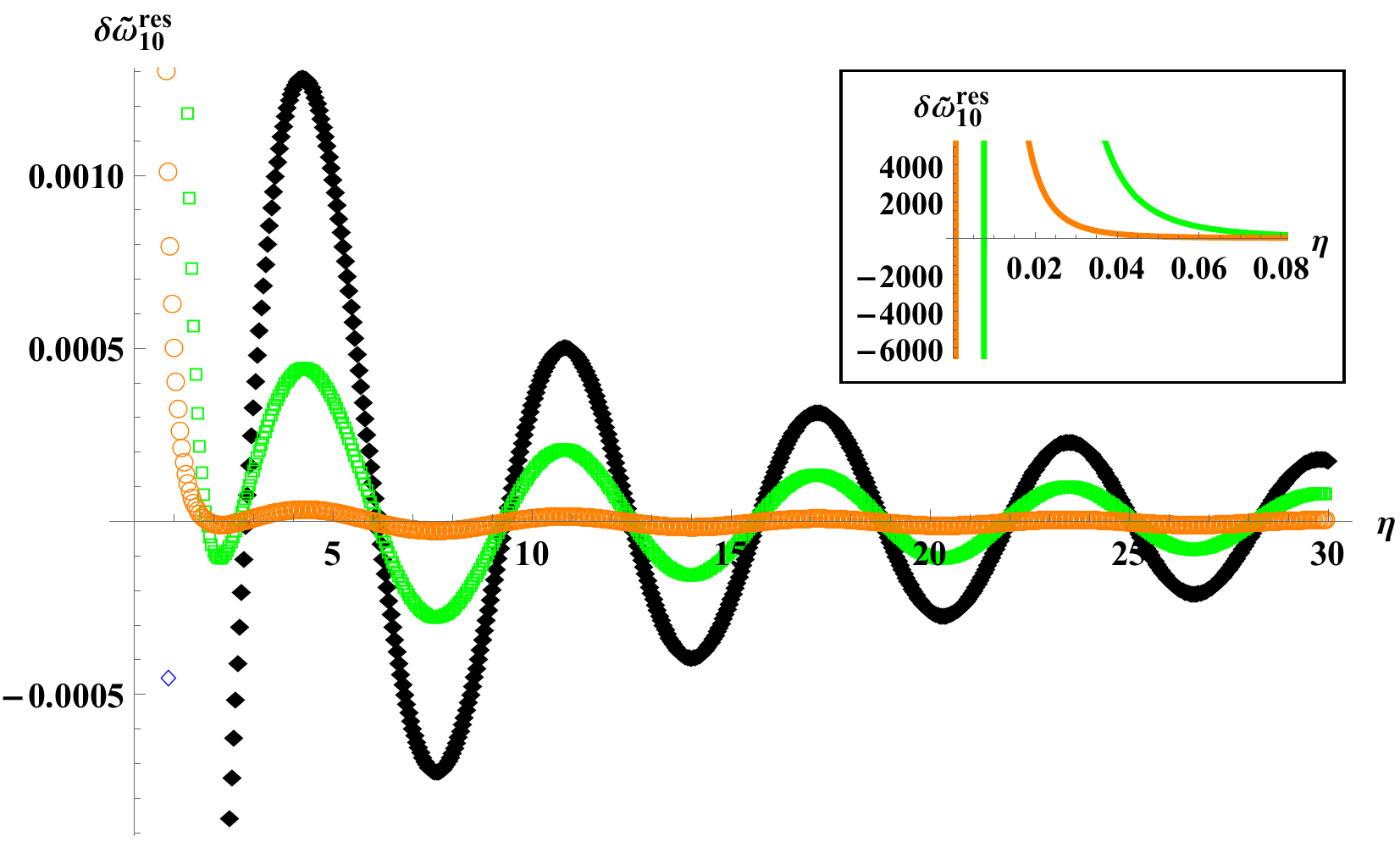}
    \caption{Resonant Casimir-Polder shift $\delta\widetilde{\omega}_{10}^{{\rm res}}$ (vertical axis) as a function of dimensionless atom-surface separation distance $\eta \equiv 2(\omega_{10}/c)z_0$ (horizontal axis) for an atomic dipole transition with a right circular polarization aligned parallel with the surface of a $C = -1$ Chern insulator, and $u/t = -1$. We show behaviors both corresponding to the case of nondispersive conductivity (black, filled diamond) as well as dispersive conductivity with $\hbar\omega_{10}/t = 10$ (green, square) and $\hbar\omega_{10}/t = 100$ (orange, circle). Top inset: Behavior of $\delta\widetilde{\omega}_{10}^{{\rm res}}$ in the near-field regime ($\eta \ll 1$) for $\hbar \omega_{10}/t = 10$ (green, square) and $100$ (orange, circle).}
   \label{resonantCPcircm2}
\end{figure}
Let us first consider the resonant Casimir-Polder shift for a right circularly polarized dipole transition. For the geometry, we can choose the $z$-axis to coincide with the normal vector to the surface and the $x$ and $y$ axes to lie in the plane of the Chern insulator. Correspondingly, $\nv^{{\rm T}} = -(1,i,0)/\sqrt{2}$, which implies that ${\bm\mu}^{10} = -\mu (1,i,0)/\sqrt{2}$ and ${\bm\mu}^{01} = -\mu (1,-i,0)/\sqrt{2}$. Since $G_{xx}^R = G_{yy}^R$  and $G_{yx}^R = -G_{xy}^R$, the use of Eq.~(\ref{formular}) leads to  
\be
\delta\widetilde{\omega}_{10}^{{\rm res}} 
=
- \frac{3}{4} \Big( {{\rm Re}}\, \widetilde{G}_{xx}^R(\rv_0,\rv_0; \tomega_{10}) 
+ {{\rm Im}}\, \widetilde{G}_{xy}^R(\rv_0,\rv_0; \tomega_{10}) \Big) \,, 
\label{E1cl-circ}
\ee
where $\widetilde{G}_{xy}^R$ is defined in Eq.~(\ref{Gxy}).  

\subsubsection{Nondispersive conductivity}

For the case where the conductivity tensor is nondispersive, $r_{sp} = r_{ps} = -C \alpha/(1 + (C \alpha)^2)$~\cite{lu2020}. 
The corresponding resonant Casimir-Polder shift experienced by a right circularly polarized dipole in the retarded regime is  
\ba 
\delta\widetilde{\omega}_{10}^{{\rm res}} 
&\!\!=\!\!& 
-\frac{3}{4} \ r_{ss} \ {{\rm Re}} \Bigg(
  \frac{i}{2} \int_{0}^{\infty} d \tk_{\parallel} \Big(\frac{\tk_{\parallel}}{\tk_z} + \tk_{\parallel}\tk_z \Big) e^{i\tk_z\eta} 
\Bigg) 
\nonumber \\
&& 
-\frac{3}{4}\ r_{sp} \ {{\rm Im}} 
\Bigg(
  i \int_{0}^{\infty} d \tk_{\parallel} \ \tk_{\parallel} \ e^{i\tk_z\eta} 
\Bigg) 
\nonumber \\
&\!\!=\!\!& 
\frac{3}{4} \Bigg( 
  \frac{(C \alpha)^2}{1 + (C \alpha)^2} \frac{(\eta^2-1)\cos \eta - \eta \sin \eta}{\eta^3} 
\nonumber\\
&& 
\quad + \ \frac{C \alpha}{1 + (C \alpha)^2} \frac{\cos \eta + \eta \sin \eta}{\eta^2} 
\Bigg) \,. 
\ea
In the far-field limit ($\eta \gg 1$), $\delta\widetilde{\omega}_{10}^{{\rm res}}$ can be approximated to leading order by 
\be
\delta\widetilde{\omega}_{10}^{{\rm res}} \approx \frac{3 C \alpha}{4(1 + (C \alpha)^2) \eta} \big(C\alpha\cos \eta + \sin \eta \big).
\ee
The resonant Casimir-Polder shift is proportional to $C\alpha$, which implies that the far-field decay behavior corresponding to a given Chern number $C$ oscillates antiphasally to
that with Chern number $-C$.   
\bing{Our far-field result in the nondispersive limit agrees with Eq.~(65) of Ref.~\cite{fuchs2017}, which is the far field formula for a circularly polarized dipole near a nondispersive Hall conductor.} 

In the near-field regime ($\eta \ll 1$), the behavior of the resonant Casimir-Polder shift can be approximated by
\be 
\delta\widetilde{\omega}_{10}^{{\rm res}} 
\approx 
- \frac{3(C \alpha)^2}{4(1 + (C \alpha)^2)\eta^3} \,. 
\label{circnonretard}
\ee
The result implies that the near-field behavior is the same for Chern numbers of the same magnitude and opposite signs.  
From the bottom right inset of Fig.~\ref{resonantCPcirc1}, we do see such a convergence of the Casimir-Polder behaviors for $C = 1$ and $C = -1$ at sufficiently small values of $\eta$. 

\begin{figure}[h]
  \centering
\includegraphics[width=0.5\textwidth]{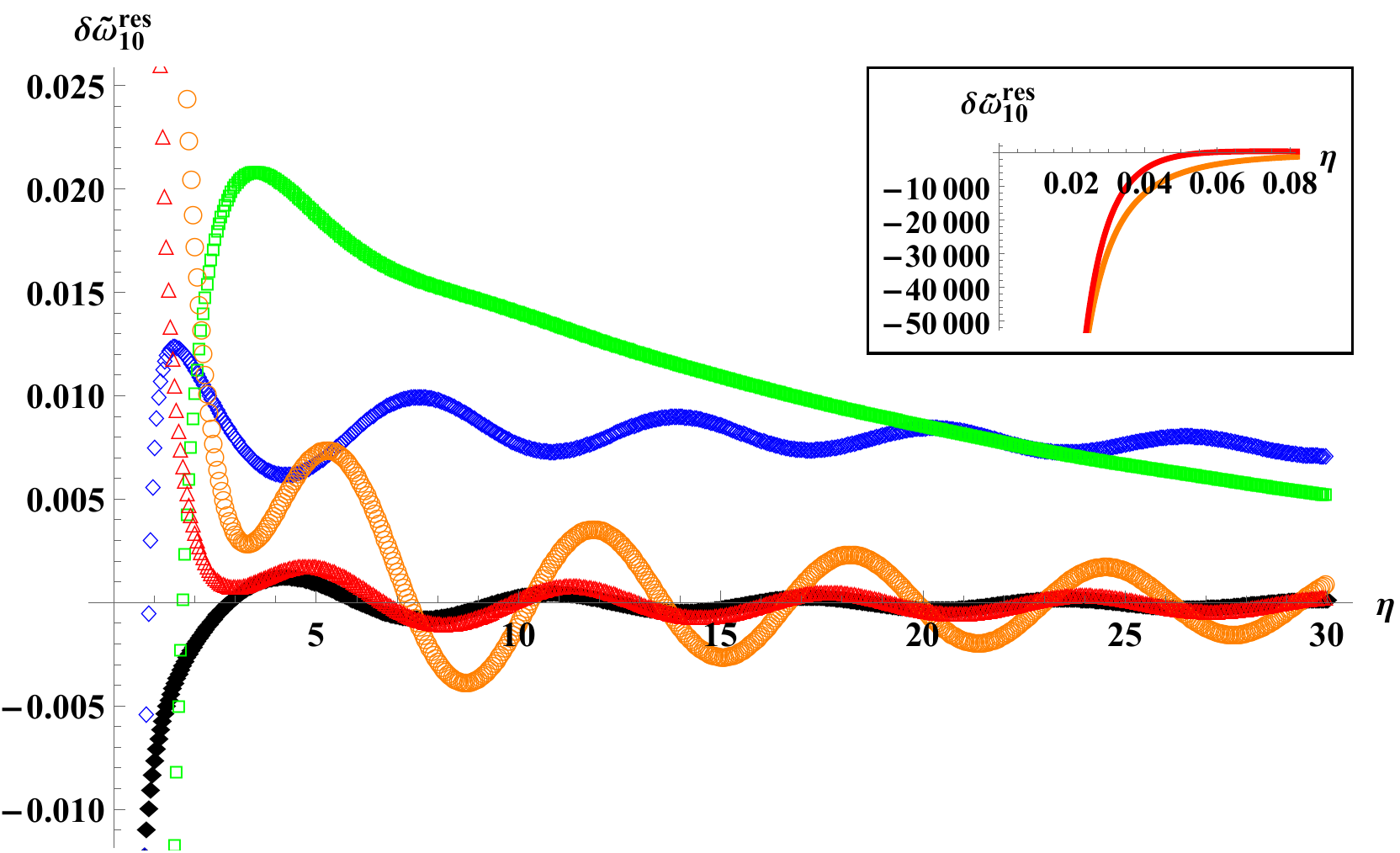}
\caption{Resonant Casimir-Polder shift $\delta\widetilde{\omega}_{10}^{{\rm res}}$ (vertical axis) as a function of dimensionless atom-surface separation distance $\eta \equiv 2(\omega_{10}/c)z_0$ (horizontal axis) for a right circularly polarized dipole located above the surface of a $C = -1$ Chern insulator with $u/t = -1$. Legend: ~nondispersive (black, filled diamond), $\hbar \omega_{10}/t = 1$ (blue, diamond), $\hbar \omega_{10}/t = 1.9$ (green, square), $\hbar \omega_{10}/t = 2.1$ (orange, circle), and $\hbar \omega_{10}/t = 3$ (red, triangle). Inset: Behavior of $\delta\widetilde{\omega}_{10}^{{\rm res}}$ in the near-field regime ($\eta \ll 1$) for the cases $\hbar \omega_{10}/t = 2.1$ and $3$.}
   \label{resonantCPcircm1}
  \end{figure} 
  
\begin{figure}[h]
  \centering
\includegraphics[width=0.45\textwidth]{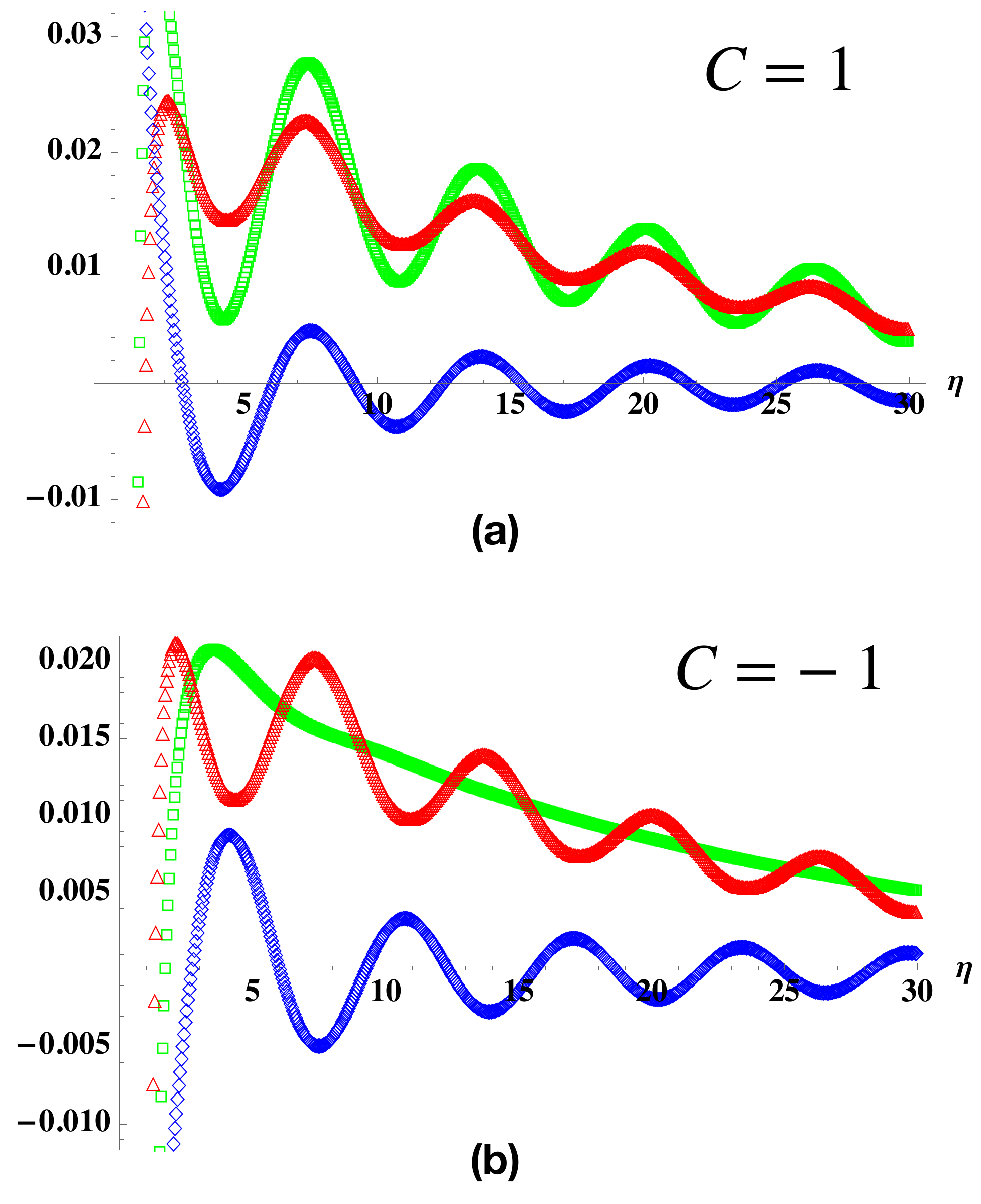}
\caption{Plots of $- (3/4) {{\rm Re}}\, \widetilde{G}_{xx}^R(\rv_0,\rv_0; \tomega_{10})$ (red triangles), $- (3/4) {{\rm Im}}\, \widetilde{G}_{xy}^R(\rv_0,\rv_0; \tomega_{10})$ (blue diamonds), and $\delta\widetilde{\omega}_{10}^{{\rm res}} =
- \frac{3}{4} ( {{\rm Re}}\, \widetilde{G}_{xx}^R(\rv_0,\rv_0; \tomega_{10}) 
+ {{\rm Im}}\, \widetilde{G}_{xy}^R(\rv_0,\rv_0; \tomega_{10}) )$ (green squares) for $\hbar\omega/t = 1.9$ and (a)~$C = 1$ and (b)~$C = -1$ Chern insulator.}
   \label{antiphase}
  \end{figure} 

\subsubsection{Effect of conductivity dispersion}
  
The inclusion of frequency dispersion into the conductivity tensor leads to a few qualitative modifications in the behavior of the resonant Casimir-Polder shift. 

The first effect that we see is an amplification of resonant Casimir-Polder shift for values of $\hbar\omega_{10}/t$ which are close to the frequency values associated with van Hove singularities. 
These singularities were found to occur at $\hbar\omega_{10} = 2t$ and $\hbar\omega_{10} = 6t$ for the case of $|u| = t$~\cite{lu2020}. 
We see the amplification in Figs.~\ref{resonantCPcirc1} and \ref{resonantCPcircm1}, which show the resonant Casimir-Polder shifts in the low and intermediate frequency regimes for a right circularly-polarized dipole transition near Chern insulators with $C=1$ and $C=-1$ respectively. 
For example, at $\eta = 1.7$ and $C=1$, the amplitude of the resonant Casimir-Polder shift for the case $\hbar \omega_{10}/t=1.9$ is 3 times larger than that for the case $\hbar \omega_{10}/t=1$, whilst the resonant Casimir-Polder shift for the case $\hbar \omega_{10}/t=2.1$ is 5 times larger than that for the case $\hbar \omega_{10}/t=3$. 
An analogous van Hove singularity-associated amplification was found for the surface correction to the atomic transition rate in Ref.~\cite{lu2020}. 
For a configuration with $C = 1$ and a right circularly polarized atomic dipole transition we see (Fig.~\ref{resonantCPcirc1}) that the resonant Casimir-Polder shift also decays with oscillations when conductivity dispersion is included, similar to the case of a nondispersive conductivity, so the force also oscillates with distance between attraction and repulsion. However, the attraction and repulsion are now much stronger than in the case of a nondispersive conductivity, for de-excitation frequencies close to the values associated with van Hove singularities. We will see in subsequent subsections that a similar behavior is exhibited in the case of an atomic dipole transition linearly polarized perpendicular or parallel to the surface. 
\bing{As we explicitly illustrate in Appendix~\ref{sec:app-extra} for the case of a circularly polarized dipole  near a $C = -1$ Chern insulator with $\widetilde{\omega}_{10} = 1.9$, the amplification of $\delta\widetilde{\omega}_{10}^{{\rm res}}$ is directly related to the possibility of enhanced (and nonzero) values of the Hall and longitudinal conductances when dispersion is taken into account, leading to the enhancement of the reflection coefficients and thus also $\delta\widetilde{\omega}_{10}^{{\rm res}}$.}

Perhaps the more interesting effect of conductivity dispersion is the possibility that the resonant Casimir-Polder shift can undergo a monotonic, non-oscillatory decay, and this possibility arises if the atomic dipole transition is circularly polarized and the insulator surface has a Hall conductance. 
From Fig.~\ref{resonantCPcircm1}, we see that such a decay occurs at distances $\eta > 5$ for $C = -1$ and a right circularly polarized atomic dipole transition at a de-excitation frequency $\omega_{10} = 1.9 t/\hbar$. 
As the corresponding force is proportional to the gradient of the Casimir-Polder shift, this implies that the force acting on the excited atom can become consistently repulsive over a much larger range of distances, compared to the case where the resonant Casimir-Polder shift undergoes an oscillatory decay.  

As the resonant Casimir-Polder shift (\ref{E1cl-circ}) gives the same value for a configuration with a right circularly polarized dipole transition and $C=-1$ and a configuration with a left circularly polarized dipole transition and $C=1$, there would also be an identical long-range repulsive force for the latter configuration. 

To understand why a monotonic, non-oscillatory decay of the resonant Casimir-Polder shift can occur, let us consider Fig.~\ref{antiphase}b, which shows the behavior corresponding to a right circularly polarized dipole transition and $C=-1$. We see that the absence of oscillations comes about because the imaginary part of $\widetilde{G}_{xy}^R$ (shown as the curve with blue diamonds) oscillates antiphasally to the real part of $\widetilde{G}_{xx}^R$ (shown as the curve with red triangles) in Eq.~(\ref{E1cl-circ}), and the oscillations from each of the terms cancel each other out to produce an overall non-oscillatory curve (shown as the curve with green squares). However, for the $C=1$ case (shown in Fig.~\ref{antiphase}a), the imaginary part of $\widetilde{G}_{xy}^R$ and the real part of $\widetilde{G}_{xx}^R$ are in phase with each other and thus the oscillations constructively interfere to produce oscillations with a more pronounced amplitude; this explains the oscillatory behavior that we see in Fig.~\ref{resonantCPcirc1}. The imaginary part of $\widetilde{G}_{xy}^R$ is proportional to $\widetilde{\sigma}_{xy}$, whose values of $C=1$ and $C=-1$ differ only by a factor of $-1$ when $\hbar\omega_{10}/t=1.9$, and this introduces phase differences of 0 and $\pi$ with respect to the real part of $\widetilde{G}_{xx}^R$ for $C=1$ and $C=-1$ cases respectively. 

Next, we consider the asymptotic behavior of the resonant Casimir-Polder shift. For all frequency regimes, we find (App.~\ref{sec:app3}) that the near-field limiting behavior of the resonant Casimir-Polder shift is given by 
\be
\delta\widetilde{\omega}_{10}^{{\rm res}} 
\approx 
-\frac{3}{4\eta^3} \,. 
\ee
We now turn to the resonant Casimir-Polder shifts for the high frequency regime ({\it i.e.}, $\hbar\omega_{10} > 6t$). The behavior is shown in Figs.~\ref{resonantCPcirc2} and \ref{resonantCPcircm2} for an excited atom with a right-circularly polarized transition near a $C=1$ Chern insulator and the same atom near a $C=-1$ Chern insulator respectively. 
We see that the resonant Casimir-Polder shift has a smaller amplitude for larger values of $\hbar\omega_{10}/t$. For example, for both $C = 1$ and $C = -1$, the amplitude of the resonant Casimir-Polder shift for $\hbar\omega_{10}/t = 100$ at $\eta = 4$ is more than 10 times smaller than that for $\hbar\omega_{10}/t = 10$. 

For all frequency regimes, the far-field behavior of the resonant Casimir-Polder shift is given by (App.~\ref{app:farfield}) 
\be
\delta\widetilde{\omega}_{10}^{{\rm res}} 
\approx 
\frac{3 \cos \eta}{8 \eta}. 
\ee
This coincides with the far-field limit of the resonant Casimir-Polder shift for the case of an atomic dipole transition aligned parallel to the surface (cf. Eq.~(\ref{farfieldresonantpara})).

\subsection{dipole polarized perpendicular to the surface}

\begin{figure}[h]
  \centering
\includegraphics[width=0.45\textwidth]{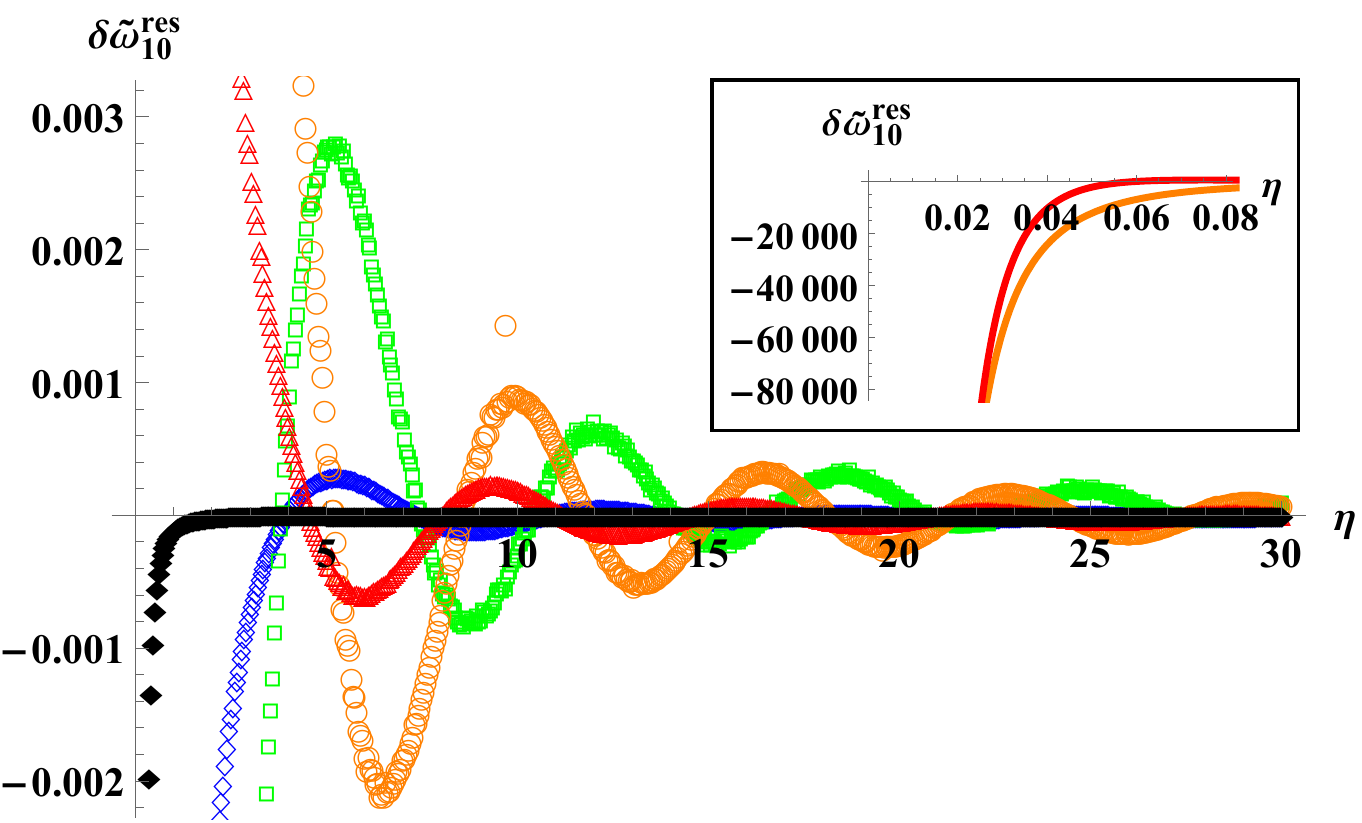}
\caption{Resonant Casimir-Polder shift $\delta\widetilde{\omega}_{10}^{{\rm res}}$ (vertical axis) as a function of dimensionless atom-surface separation distance $\eta \equiv 2(\omega_{10}/c)z_0$ (horizontal axis) for an atomic dipole transition polarized perpendicular to the surface of a $C = 1$ Chern insulator with $u/t = 1$. Legend: ~nondispersive (black, filled diamond), $\hbar \omega_{10}/t = 1$ (blue, diamond), $\hbar \omega_{10}/t = 1.9$ (green, square), $\hbar \omega_{10}/t = 2.1$ (orange, circle), and $\hbar \omega_{10}/t = 3$ (red, triangle). Top inset: Behavior of $\delta\widetilde{\omega}_{10}^{{\rm res}}$ in the near-field regime ($\eta \ll 1$) for the cases $\hbar \omega_{10}/t = 2.1$ and 3.}
\label{resonantCPvert1}
\end{figure} 
\begin{figure}[h]
  \centering
    \includegraphics[width=0.45\textwidth]{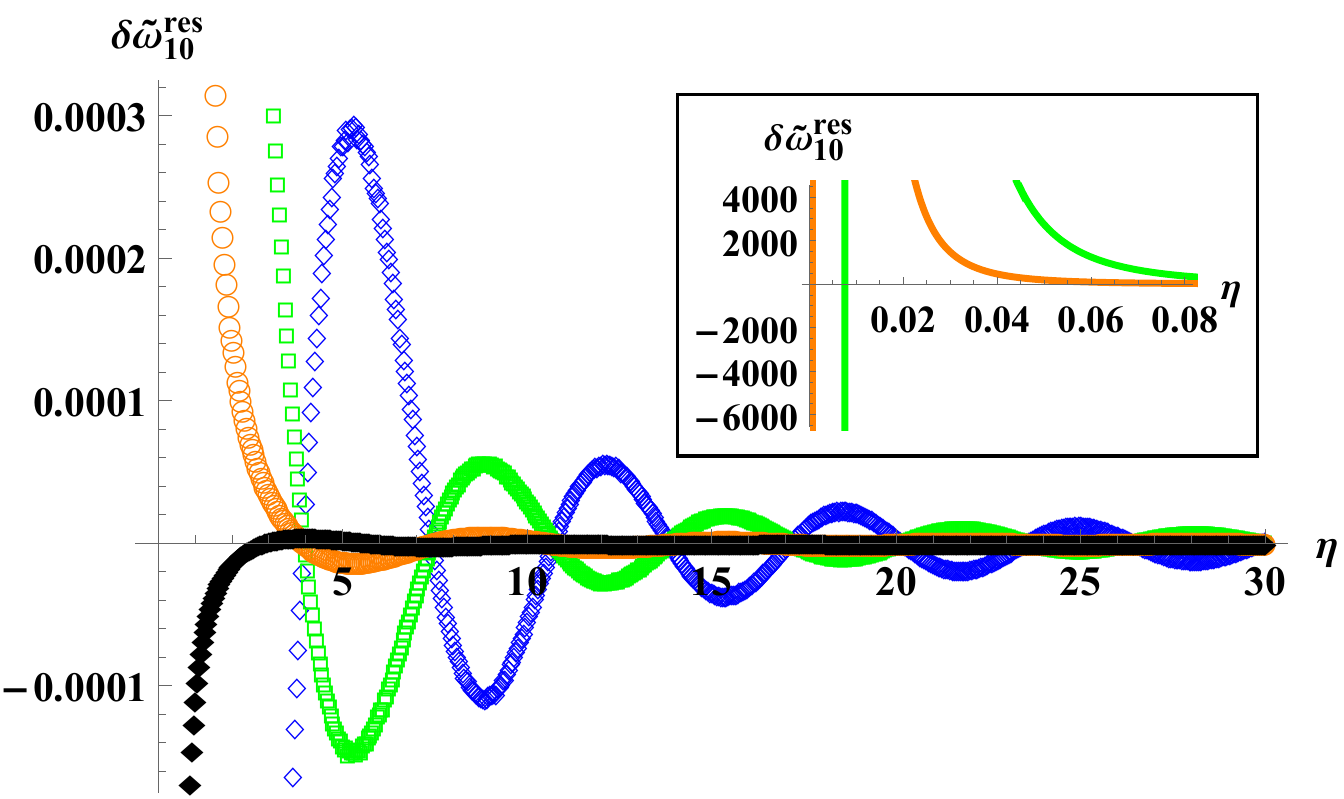}
    \caption{Resonant Casimir-Polder shift $\delta\widetilde{\omega}_{10}^{{\rm res}}$ (vertical axis) as a function of dimensionless atom-surface separation distance $\eta \equiv 2(\omega_{10}/c)z_0$ (horizontal axis) for a dipole aligned perpendicular to the surface of a $C = 1$ Chern insulator with $u/t = 1$. We show behaviors both corresponding to the case of nondispersive conductivity (black, filled diamond) as well as dispersive conductivity with $\hbar\omega_{10}/t = 1$ (blue, diamond), $\hbar\omega_{10}/t = 10$ (green, square), and $\hbar\omega_{10}/t = 100$ (orange, circle). The behavior is unchanged for $u/t = -1$ which corresponds to $C = -1$. Top inset: Behavior of $\delta\widetilde{\omega}_{10}^{{\rm res}}$ in the near-field regime ($\eta \ll 1$) for the cases $\hbar \omega_{10}/t = 10$ and 100.}
   \label{resonantCPvert2}
\end{figure}
Next, we consider an atomic dipole transition linearly polarized perpendicular to the surface of the Chern insulator, so $\nv^{{\rm T}} = (0,0,1)$. From Eq.~(\ref{formular}), we obtain 
\be
\delta\widetilde{\omega}_{10}^{{\rm res}} =
- \frac{3}{4} {{\rm Re}} \, \widetilde{G}_{zz}^R(\rv_0,\rv_0; \omega_{10}) \,,
\label{E1cl-perp}
\ee
where $\widetilde{G}_{zz}^R$ is defined in Eq.~(\ref{Gzz}).

\subsubsection{Nondispersive conductivity}
As before, we start by considering the nondispersive limit of the conductivity tensor. 
For this case, $r_{pp} = (C \alpha)^2/(1 + (C \alpha)^2)$~\cite{lu2020}, and the use of Eqs.~(\ref{Gzz}) leads to the following energy shift in the retarded regime 
\ba 
\delta\widetilde{\omega}_{10}^{{\rm res}} 
&\!\!=\!\!& 
-\frac{3 \, r_{pp}}{4} \, {{\rm Re}} 
\Bigg(i 
\int_{0}^{\infty} \!\! d \tk_{\parallel} \ \frac{\tk_{\parallel}^3}{\tk_z} \, e^{i\tk_z\eta} 
\Bigg) 
\nonumber\\
&\!\!=\!\!& 
- \frac{3(C \alpha)^2}{2(1 + (C \alpha)^2)} \frac{\cos \eta + \eta \sin \eta}{\eta^3} \,. 
\ea
In the far-field limit ($\eta \gg 1$), $\delta\widetilde{\omega}_{10}^{{\rm res}}$ can be approximated to leading order as 
\be
\delta\widetilde{\omega}_{10}^{{\rm res}} \approx - \frac{3(C \alpha)^2 \sin\eta}{2(1 + (C \alpha)^2) \eta^2} \,. 
\label{nonresnondispFFperp}
\ee
The resonant Casimir-Polder shift undergoes sinusoidal oscillations while decaying as $-1/\eta^2$, and this behavior is similar to the interaction of a dipole antenna with its reflected field~\cite{laliotis2015}. 

In the near-field/nonretarded regime ({\it i.e.}, $\eta \ll 1$), $\delta\widetilde{\omega}_{10}^{{\rm res}}$ can be approximated to leading order as
\be 
\delta\widetilde{\omega}_{10}^{{\rm res}} \approx  -\frac{3(C \alpha)^2}{2(1 + (C \alpha)^2)\eta^3}.
\label{perpnonretard}
\ee
The behavior of the resonant Casimir-Polder shift for the case of nondispersive conductivity is described by the black diamond curves in Figs.~\ref{resonantCPvert1} and \ref{resonantCPvert2}. 

\subsubsection{Effect of conductivity dispersion}
Next, we look at the effect of conductivity dispersion on the behavior of the resonant Casimir-Polder shift. 
In Fig.~\ref{resonantCPvert1}, we show the behavior of the resonant Casimir-Polder shifts for de-excitation frequencies in the low and intermediate frequency regimes, whilst in Fig.~\ref{resonantCPvert2}, we show the behavior corresponding to the high frequency regime. 
Similar to the behavior we saw for the case of a right circularly polarized transition and $C=1$, the resonant Casimir-Polder shift also decays with oscillations, and the shifts can be much enhanced for values of $\omega_{10}$ near frequencies associated with van Hove singularities. In Fig.~\ref{resonantCPvert1}, we see that such an enhancement occurs for the de-excitation frequencies $\omega_{10}=1.9t/\hbar$ and $2.1t/\hbar$. 
For example, the peak of the resonant Casimir-Polder shift for $\hbar \omega_{10}/t=1.9$ is more than 9 times larger than that for $\hbar \omega_{10}/t=1$. 

Next, let us consider the asymptotic behavior of the resonant Casimir-Polder shift. 
In the near-field limit, we find (App.~\ref{app:nearfield}) that the asymptotic behavior of the resonant Casimir-Polder shift for all frequency regimes is given by 
\be
\delta\widetilde{\omega}_{10}^{{\rm res}} 
\approx 
-\frac{3}{2\eta^3} \,.  
\ee
The near-field limit of the resonant Casimir-Polder shift is thus negative. From Figs.~\ref{resonantCPvert1} and \ref{resonantCPvert2}, it may appear that in the intermediate and high frequency regimes the resonant Casimir-Polder shift becomes infinitely large and positive as the atom-surface distance shrinks to zero. As we see in the figure insets, this is not the case. 

From Fig.~\ref{resonantCPvert2}, we also see that the amplitude of the resonant Casimir-Polder shift decreases with an increase in the value of $\hbar\omega_{10}/t$. In the high frequency regime, $\widetilde{\sigma}_{xx}'$ and $\widetilde{\sigma}_{xy}''$ vanish, whilst $\widetilde{\sigma}_{xx}''(\omega) \sim 1/\omega$ and $\widetilde{\sigma}_{xy}'(\omega) \sim 1/\omega^2$. Consequently, the far-field decay behavior of $\delta\widetilde{\omega}_{10}^{{\rm res}}$ in the high frequency regime is given by (App.~\ref{app:farfield}) 
\be
\delta\widetilde{\omega}_{10}^{{\rm res}} \approx 
\frac{3 \widetilde{\sigma}_{xx}'' \sin \eta}{4 \eta} \,. 
\ee
The inclusion of conductivity dispersion has resulted in a change to how the resonant Casimir-Polder shift scales with distance, {\it i.e.}, as $1/\eta$ rather than $1/\eta^2$ (which was the scaling behavior for the nondispersive case).



\subsection{dipole polarized parallel with the surface} 

\begin{figure}[h]
  \centering
\includegraphics[width=0.45\textwidth]{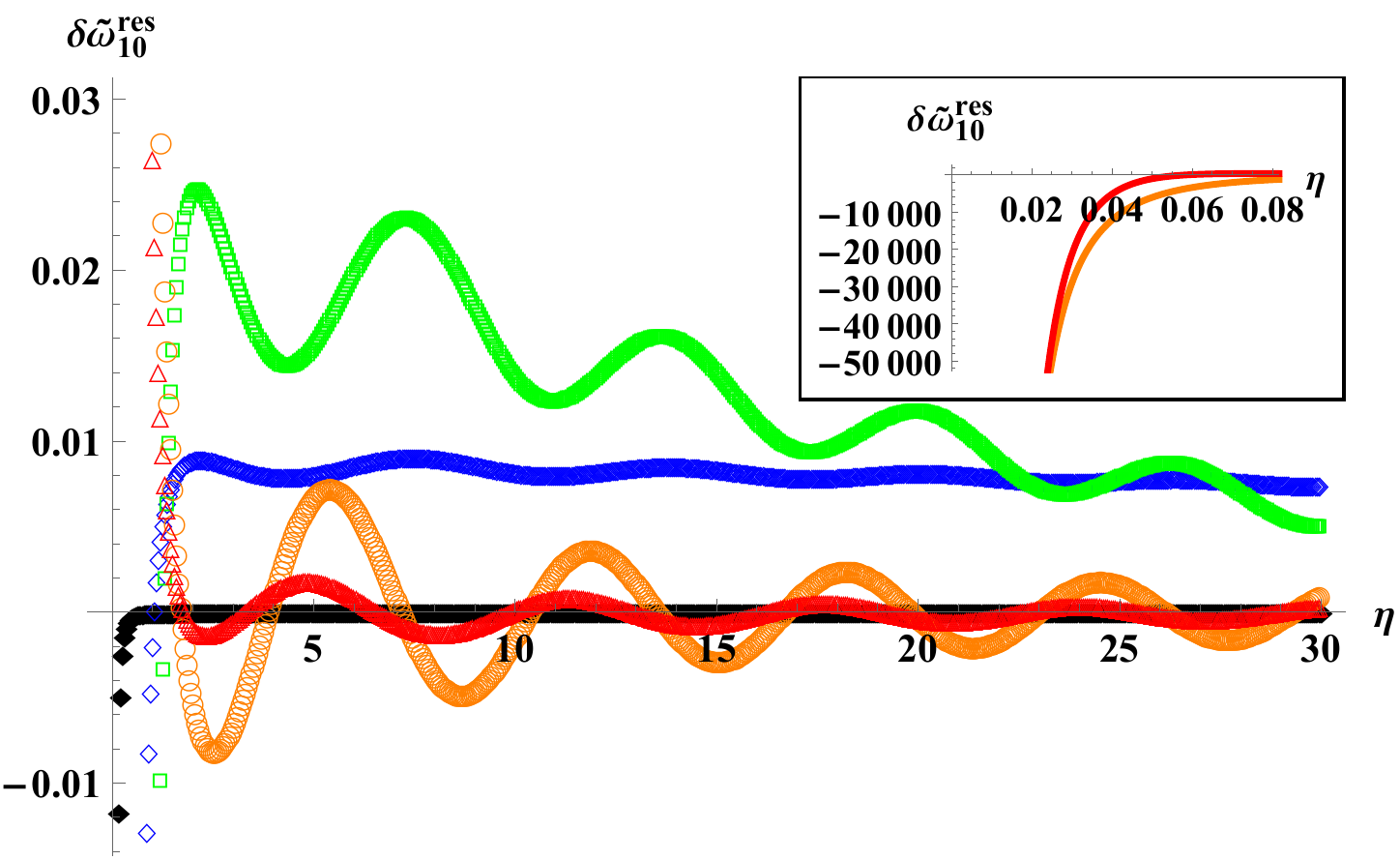}
\caption{Resonant Casimir-Polder shift $\delta\widetilde{\omega}_{10}^{{\rm res}}$ (vertical axis) as a function of dimensionless atom-surface separation distance $\eta \equiv 2(\omega_{10}/c)z_0$ (horizontal axis) for a dipole aligned parallel with the surface of a $C = 1$ Chern insulator with $u/t = 1$. Legend: ~nondispersive (black, filled diamond), $\hbar \omega_{10}/t = 1$ (blue, diamond), $\hbar \omega_{10}/t = 1.9$ (green, square), $\hbar \omega_{10}/t = 2.1$ (orange, circle), and $\hbar \omega_{10}/t = 3$ (red, triangle). Top inset: Behavior of $\delta\widetilde{\omega}_{10}^{{\rm res}}$ in the near-field regime ($\eta \ll 1$) for the cases $\hbar \omega_{10}/t = 2.1$ and 3.}
\label{resonantCPpara1}
\end{figure} 

\begin{figure}[h]
  \centering
    \includegraphics[width=0.45\textwidth]{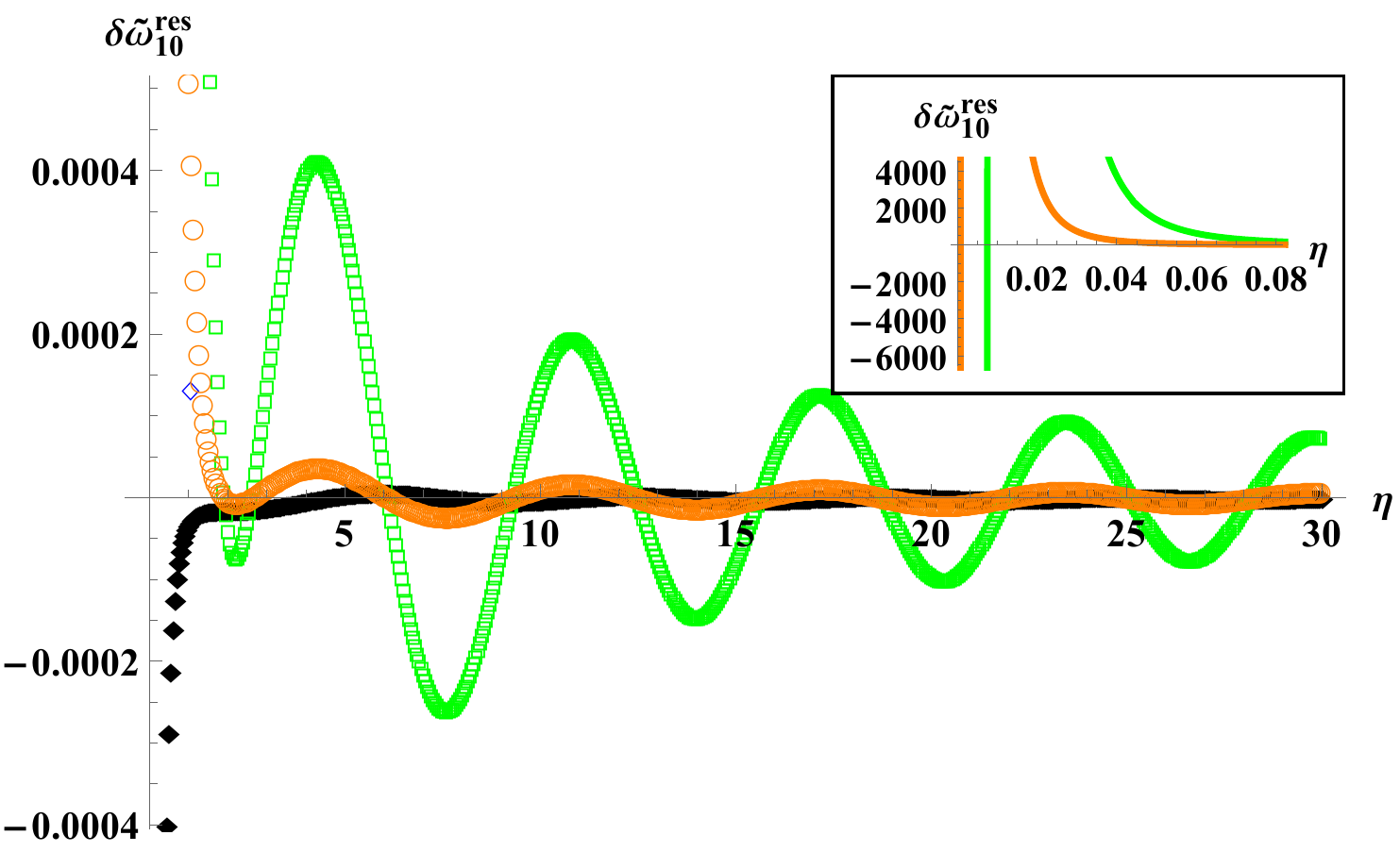}
    \caption{Resonant Casimir-Polder shift $\delta\widetilde{\omega}_{10}^{{\rm res}}$ (vertical axis) as a function of 
    dimensionless atom-surface separation distance $\eta \equiv 2(\omega_{10}/c)z_0$ (horizontal axis) for a dipole aligned parallel with the surface of a $C = 1$ Chern insulator with $u/t = 1$. We show behaviors both corresponding to the case of nondispersive conductivity (black, filled diamond) as well as dispersive conductivity with $\hbar\omega_{10}/t = 10$ (green, square) and $\hbar\omega_{10}/t = 100$ (orange, circle). The behavior is unchanged for $u/t = -1$ which corresponds to $C = -1$. Top inset: Behavior of $\delta\widetilde{\omega}_{10}^{{\rm res}}$ in the near-field regime ($\eta \ll 1$) for the cases $\hbar \omega_{10}/t = 10$ and 100.}
   \label{resonantCPpara2}
\end{figure}

Next, we study the behavior of the resonant Casimir-Polder shift for an atomic dipole transition polarized parallel to the plane of the Chern insulator. 
Without loss of generality, we choose $\nv^{{\rm T}} = (1,0,0)$. From Eq.~(\ref{formular}), we have 
\be
\delta\widetilde{\omega}_{10}^{{\rm res}} = 
- \frac{3}{4} {{\rm Re}} \, \widetilde{G}_{xx}^R(\rv_0,\rv_0; \omega_{10}) \,, 
\label{E1cl-para}
\ee
where $\widetilde{G}_{xx}^R$ is defined in Eq.~(\ref{Gxx}).  

\subsubsection{Nondispersive conductivity}
As before, we first consider a nondispersive conductivity, for which $r_{ss} = -r_{pp} = -(C \alpha)^2/(1 + (C \alpha)^2)$~\cite{lu2020}. We find that the resonant Casimir-Polder shift in the retarded regime is 
\ba 
\delta\widetilde{\omega}_{10}^{{\rm res}} 
&\!\!=\!\!& 
-\frac{3}{4} \ {{\rm Re}} 
\Bigg(
\frac{i}{2} \int_{0}^{\infty} \!\! d \tk_{\parallel} 
\Bigg[\frac{\tk_{\parallel}}{\tk_z} \ r_{ss} - \tk_{\parallel}\tk_z \ r_{pp} \Bigg] e^{i\tk_z\eta} 
\Bigg) 
\nonumber \\
&\!\!=\!\!& 
\frac{3(C \alpha)^2}{4(1 + (C \alpha)^2)} \frac{(\eta^2-1)\cos \eta - \eta \sin \eta}{\eta^3}\,. 
\ea
In the far-field limit ($\eta \gg 1$), $\delta\widetilde{\omega}_{10}^{{\rm res}}$ can be approximated to leading order as 
\be
\delta\widetilde{\omega}_{10}^{{\rm res}} 
\approx 
\frac{3(C \alpha)^2 \cos\eta}{4(1 + (C \alpha)^2) \eta} \,. 
\label{antenna-para}
\ee
The cosinusoidal decay with $1/\eta$ is reminiscent of the behavior of a classical dipole antenna aligned parallel to a surface~\cite{laliotis2015}. 

In the nonretarded regime ($\eta \ll 1$), we can approximate the frequency shift to leading order as 
\be 
\label{paranonretard}
\delta\widetilde{\omega}_{10}^{{\rm res}} \approx  - \frac{3(C \alpha)^2}{4(1 + (C \alpha)^2)\eta^3} \,.  
\ee

\subsubsection{Effect of conductivity dispersion}

Now we study the effect of conductivity dispersion. 
Figure~\ref{resonantCPpara1} shows the the behavior of the resonant Casimir-Polder shift in the low and intermediate frequency regimes. 
We see that whilst the curves for the intermediate frequency regime ({\it i.e.}, corresponding to $\hbar\omega_{10}/t = 2.1$ and 3) undergo a sinusoidal-type oscillatory decay with distance, the curves for the low frequency regime (corresponding to $\hbar\omega_{10}/t = 1$ and $1.9$) undergo a decay with sine integral-like oscillations. The latter type of oscillatory decay was  also found for the Chern insulator-induced correction to the transition rate behavior of a parallel-aligned dipole emitter in the low frequency regime~\cite{lu2020}. 
In addition, we observe that the amplitude of the resonant Casimir-Polder shift tends to decrease with larger values of $\hbar\omega_{10}/t$. For example, from Fig.~\ref{resonantCPpara2}, we see that the peak in the curve corresponding to $\hbar\omega_{10}/t = 10$ (green squares) at $\eta = 4$ is about 11 times larger than the peak for $\hbar\omega_{10}/t = 100$ (orange circles) at the same separation distance. 


In the near-field limit, the resonant Casimir-Polder shift is negative (App.~\ref{app:nearfield}) for all frequency regimes, {\it i.e.},
\be
\delta\widetilde{\omega}_{10}^{{\rm res}} 
\approx 
-\frac{3}{4\eta^3} \,.  
\ee
On the other hand, in the far-field regime, we find (App.~\ref{app:farfield}) that the resonant Casimir-Polder shift decays for all frequency regimes with the following behavior:  
\be
\delta\widetilde{\omega}_{10}^{{\rm res}} \approx \frac{3 \cos \eta}{8 \eta} \,.
\label{farfieldresonantpara} 
\ee

\section{Conclusions}

We have investigated the character of the resonant Casimir-Polder shift of an excited atom with a single (electric dipole) transition in front of a Chern insulator. 
In particular, we have studied the behavior for three different polarizations of the dipole transition of an excited atom in front of a Chern insulator with Chern number $C=1$, namely, a polarization perpendicular to the insulator surface, another polarization parallel to the surface, and finally a right circular polarization which is in the plane of the surface. For the last configuration, we have studied the behavior for both a $C=1$ and a $C=-1$ Chern insulator. For each polarization configuration, we have looked at the resonant Casimir-Polder shift for both the case where the conductivity is nondispersive and the case where the conductivity is dispersive. 
We have accounted for the frequency dispersion of the conductivity of the Chern insulator by making use of a tight-binding (Qi-Wu-Zhang) model. 

Firstly, we have found for all the configurations that the resonant Casimir-Polder shift and the corresponding force can be dramatically amplified if the de-excitation frequency of the atom is near a value associated with a van Hove singularity of the Chern insulator. We have also found that in the near-field limit, the frequency shifts diverge to negative infinity in the low, intermediate and high frequency regimes. 
In the far-field regime, for an atom whose dipole moment is right-circularly polarized,  or polarized parallel or perpendicular to a Chern insulator surface with $C = 1$, the resonant Casimir-Polder shift generally undergoes an oscillatory decay with $1/z_0$, with $z_0$ being the atom-surface separation. 


Secondly, we have also found that for a certain range of atomic de-excitation energies whose magnitudes are within the band gap of the Chern insulator, the resonant Casimir-Polder shift and the corresponding force can undergo a monotonic, non-oscillatory decay, and the force is repulsive for a much larger atom-surface separation compared to that for which the resonant Casimir-Polder shift undergoes an oscillatory decay. The monotonically decaying and repulsive force can appear in a configuration consisting of a right circularly polarized atomic dipole transition and a $C=-1$ Chern insulator, and also in a configuration consisting of a left circularly polarized atomic dipole transition and a $C=1$ Chern insulator. On the other hand, for a configuration consisting of a right circularly polarized atomic dipole transition and a $C=1$ Chern insulator or a left circularly polarized atomic dipole transition and a $C=-1$ Chern insulator, and also for a configuration involving an atomic transition linearly polarized perpendicular to or parallel to the Chern insulator surface, the resonant Casimir-Polder shift decays with oscillations, and the force also oscillates (with distance) between attraction and repulsion. 

The above finding indicates a possible mechanism for generating long-range repulsive resonant Casimir-Polder forces. 
Furthermore, the quite different behaviors of the resonant Casimir-Polder shift can provide a way to distinguish the sign of the Chern number for a given circular polarization of the atom, and conversely also enable one to distinguish the direction of an atom's circular polarization for a given sign of the Chern number. 
The ideas and methodology described in this paper can be straightforwardly generalized to the case of atoms with more than one transition and Chern insulators with higher Chern numbers. 

\section{Acknowledgments}

One of the authors (BSL) thanks David Wilkowski for constructive discussions. This work was supported by funding from the Research Development Funding of Xi'an Jiaotong-Liverpool University (XJTLU) under the grant number RDF-21-02-005. 

\section{Author contribution statement}

All authors contributed equally to the paper. 

\section{Data availability statement}

The datasets generated during and/or analysed during the current study are available from the corresponding author on reasonable request.


\appendix

\begin{widetext}

\section{Fresnel coefficients} 
\label{app:fresnel}   

In this Appendix, we present the formulas (with dimensions restored) for the Fresnel coefficients obtained for the case of a single Chern insulator~\cite{lu2020}.  
We denote by the symbols $r_{ss}$, $r_{sp}$, $r_{ps}$ and $r_{pp}$ respectively the reflection coefficients for an incident s-polarized wave which is reflected as an s-polarized wave, an incident s-polarized wave which is reflected as an p-polarized wave, an incident p-polarized wave which is reflected as an s-polarized wave, and an incident p-polarized wave which is reflected as an p-polarized wave. For the transmission coefficients, we replace the symbol $r$ by the symbol $t$. 
\begin{subequations}
\ba
r_{ss} &\!\!=\!\!& 
-\frac{1}{\Delta} \bigg( \frac{4\pi^2}{c^2} (\sigma_{xx}^2 + \sigma_{xy}^2) 
+ \frac{2\pi}{c} \big( 1 - (ck_\parallel/\omega)^2 \big)^{-1/2} \sigma_{xx} \bigg), 
\\
r_{ps} &\!\!=\!\!& r_{sp} = - \frac{2\pi}{c} \frac{\sigma_{xy}}{\Delta}, 
\\
r_{pp} &\!\!=\!\!& 
\frac{1}{\Delta} \bigg( \frac{4\pi^2}{c^2} (\sigma_{xx}^2 + \sigma_{xy}^2) 
+ \frac{2\pi}{c} \big( 1 - (ck_\parallel/\omega)^2 \big)^{1/2} \sigma_{xx} \bigg), 
\\
t_{ss} &\!\!=\!\!& 
\frac{1}{\Delta} 
\bigg( 1 + \frac{2\pi}{c} \big( 1 - (ck_\parallel/\omega)^2 \big)^{1/2} \sigma_{xx} \bigg), 
\\
t_{ps} &\!\!=\!\!& t_{sp} = - \frac{2\pi}{c} \frac{\sigma_{xy}}{\Delta}, 
\\
t_{pp} &\!\!=\!\!& 
\frac{1}{\Delta} 
\bigg( 1 + \frac{2\pi}{c} \big( 1 - (ck_\parallel/\omega)^2 \big)^{-1/2} \sigma_{xx} \bigg), 
\ea
\end{subequations}
where
$\Delta \equiv 1 + \frac{2\pi}{c} \Big( \big( 1 - (ck_\parallel/\omega)^2 \big)^{1/2} + \big( 1 - (ck_\parallel/\omega)^2 \big)^{-1/2} \Big) \sigma_{xx} + \frac{4\pi^2}{c^2} \big( \sigma_{xx}^2 + \sigma_{xy}^2 \big)$. 
By defining a dimensionless conductivity $\tsigma \equiv 2\pi \sigma/c$, we can put the reflection coefficients above in the form Eqs.~(\ref{r-coeffs}).  

\section{derivation of energy shifts in the presence of a nonreciprocal medium} 
\label{sec:app1} 

In Ref.~\cite{wylie-sipe2}, it was shown using perturbation theory methods that for a quantum emitter interacting with the radiation field via the dipole interaction $-\bm{\mu}\cdot\Dv$, the shift $\delta E_m'$ in the energy $E_m$ of atomic state $|m\rangle$ is given by 
\be
\label{A1}
\delta E_m'  = 
- \mathcal{P} 
\left\{ 
\sum_{n} 
\int_{-\infty}^\infty \!\!\! d\omega \, \frac{\mu_a^{mn} \mu_b^{nm} }{\omega + \omega_{nm}}
\left[ 
\frac{1}{\hbar} 
\sum_{B,N} p(B)  
D_a^{BN}(\rv_0) D_b^{NB}(\rv_0) \, \delta(\omega - \omega_{NB})
\right] 
\right\}.
\ee
Here, $\mathcal{P}$ denotes the principal value, $a, b = 1,2,3$ label Cartesian axes, $m, n$ label atomic states, $B, N$ label field states, $p(B)$ denotes the prior probability to prepare the field in state $|B\rangle$, $\omega_{nm} \equiv (E_n - E_m)/\hbar$, $\mu_a^{mn} \equiv \langle m | \mu_a | n \rangle$ denotes the dipole transition matrix element from atomic state $|n\rangle$ to $|m\rangle$, and $D_a^{BN} \equiv \langle B | D_a | N \rangle$ denotes the dipole transition matrix element from field state $|N\rangle$ to $|B\rangle$. 
As we are interested in the surface-induced correction to the energy shift, we have ignored a contribution which arises purely from the electromagnetic fluctuations in free space. 

In the presence of a nonreciprocal medium, it turns out that the term enclosed by the square brackets in Eq.~(\ref{A1}) is proportional to the anti-Hermitian part of the dyadic Green function. In what follows, we are going to derive this result. We first recall that the dyadic Green tensor $\mathbb{G}(\rv,\rv',t)$ connects the dipole $\pv(t')$ at time $t'$ and position $\rv'$ to the expectation value of its displacement field response, $\langle \Dv(\rv,t) \rangle$, via 
\be
\langle D_a(\rv,t) \rangle = \int_{-\infty}^t \!\!\!dt' \, G_{ab}(\rv,\rv',t-t') \, p_b(t'), 
\ee
with the dyadic Green tensor being given by~\cite{fain} 
\be
G_{ab}(\rv,\rv',t) = \frac{i}{\hbar} \sum_{B,N} p(B) \big[ \langle B | D_a(\rv,t) | N \rangle \langle N | D_b(\rv',0) |B\rangle - \langle B | D_b(\rv' , 0) | N \rangle \langle N | D_a(\rv,t) |B\rangle \big] \Theta(t), 
\ee
where $\Theta(t)$ is the Heaviside step function, being equal to 1 if $t > 0$ and 0 if $t < 0$, and the operator $D_a(\rv,t)$ is in the interaction picture. The operator in the interaction picture  is related to its counterpart $D_a(\rv)$ in the Schr\"{o}dinger picture by $D_a(\rv,t) = e^{iH_0 t/\hbar} D_a(\rv) \, e^{-iH_0 t/\hbar}$, where $H_0$ is the free Hamiltonian for the field in the absence of the dipole interaction. In terms of the Schr\"{o}dinger picture, the dyadic Green tensor is given by 
\be
G_{ab}(\rv,\rv',t) = \frac{i}{\hbar} \sum_{B,N} p(B) 
\big[ 
e^{-i\omega_{NB}t} 
D_a^{BN}(\rv) D_b^{NB}(\rv') 
- e^{i\omega_{NB}t} 
D_b^{BN}(\rv') D_a^{NB}(\rv)  
\big] 
\Theta(t). 
\ee
By relabeling $B \leftrightarrow N$ in the second term, we obtain 
\be
G_{ab}(\rv,\rv',t) = 
\frac{i}{\hbar} \sum_{B,N} 
e^{-i\omega_{NB}t} 
\big[ p(B) - p(N) \big] 
D_a^{BN}(\rv) D_b^{NB}(\rv')  
\Theta(t). 
\ee
The Fourier transform is given by 
\ba
G_{ab}(\rv,\rv',\omega) 
&\!\!=\!\!& 
\int_{-\infty}^{\infty} \!\!\! dt \, e^{i\omega t} G_{ab}(\rv,\rv',t) 
\nonumber\\
&\!\!=\!\!& 
-\frac{1}{\hbar} \sum_{B,N} 
\big[ p(B) - p(N) \big] 
\frac{D_a^{BN}(\rv) D_b^{NB}(\rv')}{\omega - \omega_{NB} + i0}
\nonumber\\
&\!\!=\!\!& 
-\frac{1}{\hbar} \sum_{B,N} 
\big[ p(B) - p(N) \big] 
D_a^{BN}(\rv) D_b^{NB}(\rv') 
\left[ 
\mathcal{P} \left( \frac{1}{\omega - \omega_{NB}} \right) - i\pi \delta(\omega - \omega_{NB})
\right]. 
\ea
As the displacement field is a physical observable, $\widehat{\Dv} = \widehat{\Dv}^\dagger$, which implies $(D_{a}^{BN})^\ast = D_{a}^{NB}$. This implies 
\be
G_{ba}^\ast(\rv',\rv,\omega) =
-\frac{1}{\hbar} 
\sum_{B,N} 
\big[ p(B) - p(N) \big] 
D_a^{BN}(\rv) D_b^{NB}(\rv') 
\left[ 
\mathcal{P} \left( \frac{1}{\omega - \omega_{NB}} \right) + i\pi \delta(\omega - \omega_{NB})
\right]. 
\ee
Noting that $p(B) - p(N) = p(B) (1 - e^{-\beta \hbar \omega_{NB}})$ and making use of the Dirac delta to change $\omega_{NB}$ to $\omega$, we obtain
\be
G_{ab}(\rv,\rv',\omega)  - G_{ba}^\ast(\rv',\rv,\omega) 
= \frac{2\pi i}{\hbar} 
\sum_{B,N} 
p(B) 
D_a^{BN}(\rv) D_b^{NB}(\rv') 
\big( 1 - e^{-\beta \hbar \omega} \big) 
\delta(\omega - \omega_{NB}), 
\ee
which implies 
\be
\frac{1}{\hbar} \sum_{B,N} 
p(B) 
D_a^{BN}(\rv) D_b^{NB}(\rv') 
\delta(\omega - \omega_{NB})
=
\frac{1}{2\pi i} \frac{G_{ab}(\rv,\rv',\omega)  - G_{ba}^\ast(\rv',\rv,\omega)}{1 - e^{-\beta \hbar \omega}}. 
\ee
We can rewrite the denominator of the rightmost term in terms of the Bose-Einstein distribution function, $n(T,\omega)=(e^{\beta\hbar\omega}-1)^{-1}$ as follows
\be
\frac{1}{\hbar} \sum_{B,N} 
p(B) 
D_a^{BN}(\rv) D_b^{NB}(\rv') 
\delta(\omega - \omega_{NB})
=
-\frac{n(T,-\omega)}{2\pi i} \bigg(G_{ab}(\rv,\rv',\omega)  - G_{ba}^\ast(\rv',\rv,\omega)\bigg).
\ee
When the temperature of the system is at or close to $T=0K$, the Bose-Einstein distribution function can be approximated in terms of Heaviside function as $n(T,\omega)\to -\Theta(-\omega)$, and hence we have that $n(T,-\omega)\to -\Theta(\omega)$:
\be
\frac{1}{\hbar} \sum_{B,N} 
p(B) 
D_a^{BN}(\rv) D_b^{NB}(\rv') 
\delta(\omega - \omega_{NB})
=
\frac{\Theta(\omega)}{2\pi i} \bigg(G_{ab}(\rv,\rv',\omega)  - G_{ba}^\ast(\rv',\rv,\omega)\bigg).
\ee
Using this expression, we can rewrite the energy shift $\delta E'_m$ in Eq.~(\ref{A1}) as
\ba
\delta E_m' 
&\!\!=\!\!& -\frac{1}{2\pi i} \mathcal{P} \left( \sum_n \mu_{a}^{mn} \mu_{b}^{nm} \!\! \int_{-\infty}^{\infty} \!\!\!\! d\omega \, \frac{\Theta(\omega)\big(G_{ab}(\rv_0,\rv_0; \omega) - G_{ba}^\ast(\rv_0,\rv_0; \omega)\big)}{\omega + \omega_{nm}} \right). 
\ea
Since the Heaviside function $\Theta(\omega)$ has values of 1 and 0 if its argument $\omega$ takes on positive and negative values respectively, only the positive frequency integral will survive. Hence we get
\ba
\delta E_m' 
&\!\!=\!\!& -\frac{1}{2\pi i} \mathcal{P} \left( \sum_n \mu_{a}^{mn} \mu_{b}^{nm} \!\! \int_0^\infty \!\!\!\! d\omega \frac{G_{ab}(\rv_0,\rv_0; \omega) - G_{ba}^\ast(\rv_0,\rv_0; \omega)}{\omega + \omega_{nm}} \right) 
\nonumber\\
&\!\!=\!\!&
-\frac{1}{2\pi i} \sum_n 
\mu_{a}^{mn} \mu_{b}^{nm} 
\!\! \int_0^\infty \!\!\!\! d\omega 
\frac{G_{ab}(\rv_0,\rv_0; \omega) - G_{ba}^\ast(\rv_0,\rv_0; \omega)}{\omega + \omega_{nm} + i\eta}
\nonumber\\
&&-\frac{1}{2} \sum_n 
\mu_{a}^{mn} \mu_{b}^{nm} 
\!\! \int_0^\infty \!\!\!\! d\omega 
\big(
G_{ab}(\rv_0,\rv_0; \omega) - G_{ba}^\ast(\rv_0,\rv_0; \omega)
\big)
\delta(\omega + \omega_{nm}). 
\ea
In the above, on going to the second step, we have made use of the Plemelj relation, $\mathcal{P}(1/x) = 1/(x+i\eta) + i\pi\delta(x)$. Next we make use of the Schwarz reflection property $G_{ba}^\ast(\omega) = G_{ba}(-\omega^\ast)$ (which is simply a statement about the reality of the Green function in real time domain) to obtain 
\ba
\label{Em'}
\delta E_m' 
&\!\!=\!\!& 
-\frac{1}{2\pi i} \sum_n 
\mu_{a}^{mn} \mu_{b}^{nm} 
\!\! \int_0^\infty \!\!\!\! d\omega 
\frac{G_{ab}(\rv_0,\rv_0; \omega)}{\omega - \omega_{mn} + i\eta}
-\frac{1}{2\pi i} \sum_n 
\mu_{a}^{mn} \mu_{b}^{nm} 
\!\! \int_{-\infty}^0 \!\!\!\! d\omega 
\frac{G_{ba}(\rv_0,\rv_0; \omega)}{\omega + \omega_{mn} - i\eta}
\nonumber\\
&&-\frac{1}{2} \sum_n 
\mu_{a}^{mn} \mu_{b}^{nm} 
\!\! \int_0^\infty \!\!\!\! d\omega \, 
\delta(\omega - \omega_{mn}) G_{ab}(\rv_0,\rv_0; \omega)
+\frac{1}{2} \sum_n 
\mu_{a}^{mn} \mu_{b}^{nm} 
\!\! \int_{-\infty}^0 \!\!\!\! d\omega \, 
\delta(\omega + \omega_{mn}) G_{ba}(\rv_0,\rv_0; \omega)
\ea
\begin{figure}[h]
\centering
  \includegraphics[width=0.46\textwidth]{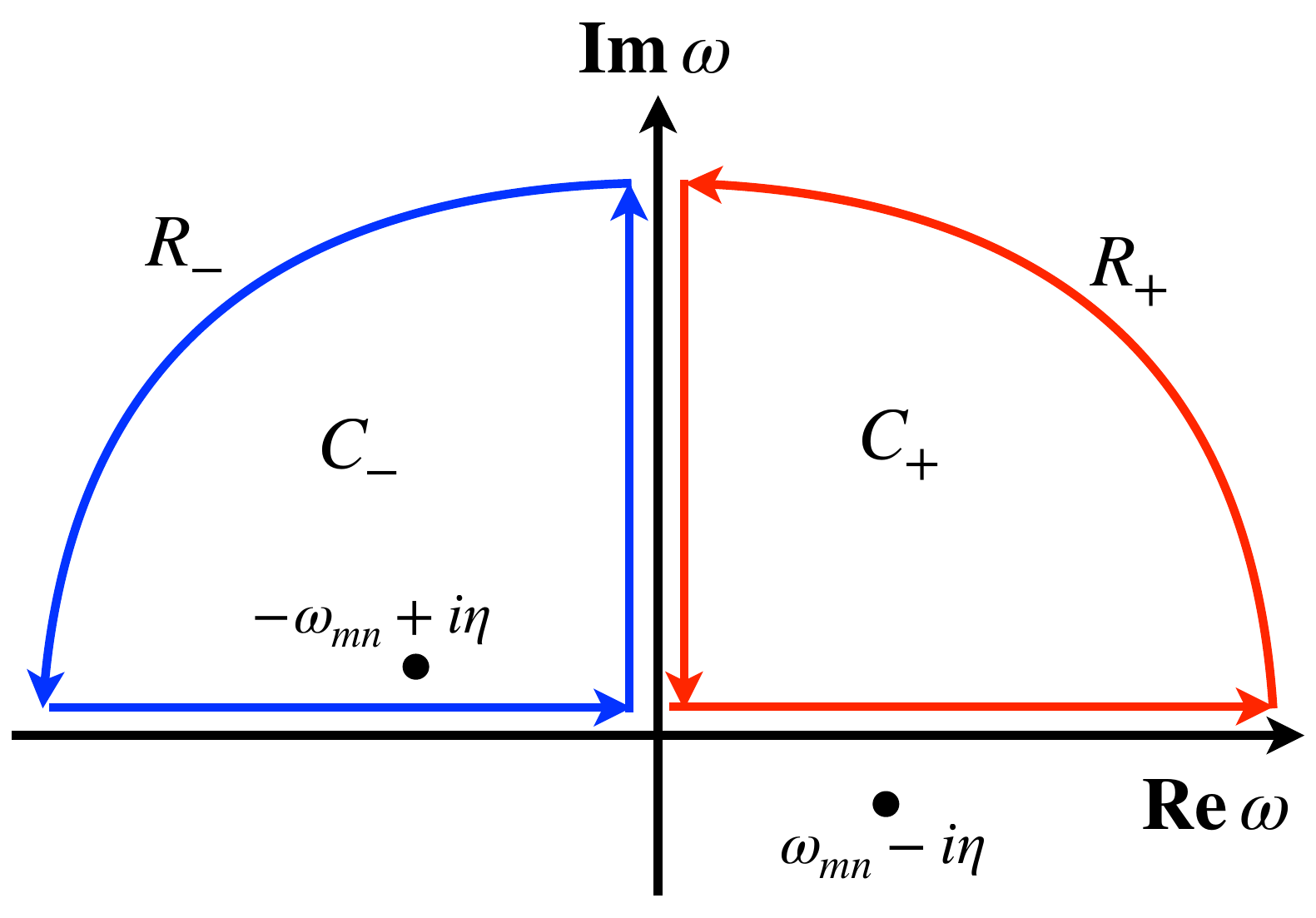}
  \caption{Integration contours on the complex frequency plane. There are two poles, which are located at $\omega = \pm(\omega_{mn} - i\eta)$.} 
  \label{contour}
\end{figure}
To evaluate the first term, we consider integrating around the contour $C_+$ in Fig.~\ref{contour}, which does not enclose any pole, and the segment $R_+$ tends radially to infinity. By the residue theorem we have 
\ba
&&\oint_{C_+} \frac{G_{ab}(\rv_0,\rv_0; \omega) d\omega}{\omega - \omega_{mn} + i\eta} = 
\Big( \int_0^\infty + \int_{R_+} - \int_0^{i\infty} \Big)
 \frac{G_{ab}(\rv_0,\rv_0; \omega) d\omega}{\omega - \omega_{mn} + i\eta} 
 = 0. 
\ea 
By Jordan's Lemma, the integral over $R_+$ vanishes as $|\omega| \to \infty$ for $G_{ab}$ becomes exponentially suppressed. Thus, 
\be
\label{plusinfty}
\int_0^\infty \!\! \frac{G_{ab}(\rv_0,\rv_0; \omega) \, d\omega}{\omega - \omega_{mn} + i\eta} 
= i \int_0^\infty \!\! \frac{G_{ab}(\rv_0,\rv_0; i\xi) \, d\xi}{- \omega_{mn} + i \xi + i\eta}. 
\ee
Next, to evaluate the second term in Eq.~(\ref{Em'}), we consider integrating around contour $C_-$ in Fig.~\ref{contour}, which encloses a pole at $\omega = -\omega_{mn} + i\eta$, and the segment $R_-$ tends radially to infinity. By the residue theorem we have 
\ba
\oint_{C_-} \frac{G_{ba}(\rv_0,\rv_0; \omega) \, d\omega}{\omega + \omega_{mn} - i\eta} 
&=& 2\pi i \,  G_{ba}(\rv_0,\rv_0; -\omega_{mn} + i\eta) \, \Theta(\omega_{mn})
 \nonumber\\
&=& \Big( \int_{-\infty}^0 + \int_0^{i\infty} + \int_{R_-} \Big)
 \frac{G_{ba}(\rv_0,\rv_0; \omega) d\omega}{\omega + \omega_{mn} - i\eta}.
 \nonumber
\ea 
Again, Jordan's Lemma implies we can neglect the integral over $R_-$, and we thus have 
\be
\label{minusinfty}
\int_{-\infty}^0 \! \frac{G_{ba}(\rv_0,\rv_0; \omega) \, d\omega}{\omega + \omega_{mn} - i\eta} 
=
- 
i \! \int_0^{\infty} \!\! \frac{G_{ba}(\rv_0,\rv_0; i\xi) \, d\xi}{\omega_{mn} + i\xi - i\eta}
+ 2\pi i \,  G_{ba}(\rv_0,\rv_0; -\omega_{mn} + i\eta) \, \Theta(\omega_{mn}) . 
\ee 
As $\int_0^\infty d\omega \, 
\delta(\omega - \omega_{mn}) G_{ab}(\omega) = G_{ab}(\omega_{mn})\Theta(\omega_{mn})$ and $\int_{-\infty}^0 d\omega \, \delta(\omega + \omega_{mn}) G_{ba}(\omega) = G_{ba}(- \omega_{mn}) \Theta(\omega_{mn})$, and using Eqs.~(\ref{plusinfty}) and (\ref{minusinfty}), we can express Eq.~(\ref{Em'}) as 
\ba
\delta E_m' 
&\!\!=\!\!& 
\frac{1}{\pi} \sum_n \mu_a^{mn} \mu_b^{nm} \!\! 
\int_0^\infty \!\!\!\!\! d\xi 
\Big(
\frac{\omega_{mn} \big( G_{ab}(\rv_0,\rv_0; i\xi) + G_{ba}(\rv_0,\rv_0; i\xi) \big)}{2(\omega_{mn}^2 + \xi^2)}
-
\frac{\xi \big( G_{ab}(\rv_0,\rv_0; i\xi) - G_{ba}(\rv_0,\rv_0; i\xi) \big)}{2i(\omega_{mn}^2 + \xi^2)}
\Big)
\nonumber\\
&&- 
\frac{1}{2} \sum_n \mu_a^{mn} \mu_b^{nm} \Theta(\omega_{mn}) 
\big( G_{ab}(\rv_0,\rv_0; \omega_{mn}) 
+ G_{ba}(\rv_0,\rv_0; -\omega_{mn}) \big)
\nonumber\\
&\!\!=\!\!& 
\frac{1}{\pi} \sum_n \mu_a^{mn} \mu_b^{nm} \!\! 
\int_0^\infty \!\!\!\!\! d\xi 
\Big(
\frac{\omega_{mn} \big( G_{ab}(\rv_0,\rv_0; i\xi) + G_{ba}^\ast(\rv_0,\rv_0; i\xi) \big)}{2(\omega_{mn}^2 + \xi^2)}
-
\frac{\xi \big( G_{ab}(\rv_0,\rv_0; i\xi) - G_{ba}^\ast(\rv_0,\rv_0; i\xi) \big)}{2i(\omega_{mn}^2 + \xi^2)}
\Big)
\nonumber\\
&&- 
\frac{1}{2} \sum_n \mu_a^{mn} \mu_b^{nm} \Theta(\omega_{mn}) 
\big( G_{ab}(\rv_0,\rv_0; \omega_{mn}) 
+ G_{ba}^\ast(\rv_0,\rv_0; \omega_{mn}) \big).
\label{Em'fina}
\ea
On going to the second equality, we have again made use of the Schwarz reflection property: $G_{ba}^\ast(\omega) = G_{ba}(-\omega^\ast)$. For (real) imaginary frequencies, this implies $G_{ba}(-\omega) = G_{ba}^\ast(\omega)$ ($G_{ba}(i\xi) = G_{ba}^\ast(i\xi)$). Equation~(\ref{Em'fina}) generalizes the formula for the atomic energy level shift to nonreciprocal media. 

The first two terms of Eq.~(\ref{Em'fina}) represent the nonresonant contribution to the Casimir-Polder shift, whilst the third term represents the resonant contribution. 
For a two-level atom with a single (dipole) transition, the third term leads to $\delta E_1^{{\rm cl}}$ and $\delta E_0^{{\rm cl}}$ in Eq.~(\ref{marimo}). 

We can check that for reciprocal normal insulators (for which $G_{ba}^\ast(\rv_0,\rv_0; \omega) = G_{ab}^\ast(\rv_0,\rv_0; \omega)$), we recover Eq.~(2.22) of Ref.~\cite{wylie-sipe2}. Our result~(\ref{Em'fina}) also agrees with the result obtained from an atomic dynamics-based approach~\cite{fuchs-thesis}. 

\section{Green tensor contractions for various dipole configurations} 
\label{sec:app2} 

In this Appendix, we collect the results for the contractions of the reflection Green tensor with the dipole orientation vector for three different dipole configurations. 

\subsection{dipole polarization perpendicular to the surface}

For the case $\nv^{{\rm T}} = (0,0,1)$, we find 

\begin{subequations}
\label{G-identities-perp}
\ba
n_a \big( G_{ab}^R + G_{ba}^{R\ast} \big) n_b^\ast 
&\!\!\!=\!\!\!& 
2 {{\rm Re}}\, \big( G_{zz}^R \big), 
\\
n_a^\ast \big( G_{ab}^R + G_{ba}^{R\ast} \big) n_b 
&\!\!\!=\!\!\!& 
2 {{\rm Re}}\, \big( G_{zz}^R \big),
\\
n_a \big( G_{ab}^R - G_{ba}^{R\ast} \big) n_b^\ast 
&\!\!\!=\!\!\!& 
2i {{\rm Im}}\, \big( G_{zz}^R \big),
\\
n_a^\ast \big( G_{ab}^R - G_{ba}^{R\ast} \big) n_b 
&\!\!\!=\!\!\!& 
2i {{\rm Im}}\, \big( G_{zz}^R \big).  
\ea
\end{subequations}

\subsection{dipole polarization parallel to the surface}

For the case $\nv^{{\rm T}} = (1,0,0)$, we find 

\begin{subequations}
\label{G-identities-para}
\ba
n_a \big( G_{ab}^R + G_{ba}^{R\ast} \big) n_b^\ast 
&\!\!\!=\!\!\!& 
2 {{\rm Re}}\, \big( G_{xx}^R \big), 
\\
n_a^\ast \big( G_{ab}^R + G_{ba}^{R\ast} \big) n_b 
&\!\!\!=\!\!\!& 
2 {{\rm Re}}\, \big( G_{xx}^R \big),
\\
n_a \big( G_{ab}^R - G_{ba}^{R\ast} \big) n_b^\ast 
&\!\!\!=\!\!\!& 
2i {{\rm Im}}\, \big( G_{xx}^R \big),
\\
n_a^\ast \big( G_{ab}^R - G_{ba}^{R\ast} \big) n_b 
&\!\!\!=\!\!\!& 
2i {{\rm Im}}\, \big( G_{xx}^R \big).  
\ea
\end{subequations}

\subsection{right circular dipole polarization} 

Finally, for the case $\nv^{{\rm T}} = (1,i,0)/\sqrt{2}$, we find 
\begin{subequations}
\label{G-identities-circ}
\ba
n_a \big( G_{ab}^R + G_{ba}^{R\ast} \big) n_b^\ast 
&\!\!\!=\!\!\!& 
2 \big( {{\rm Re}}\, G_{xx}^R + {{\rm Im}}\, G_{xy}^R \big), 
\\
n_a^\ast \big( G_{ab}^R + G_{ba}^{R\ast} \big) n_b 
&\!\!\!=\!\!\!& 
2 \big( {{\rm Re}}\, G_{xx}^R - {{\rm Im}}\, G_{xy}^R \big), 
\\
n_a \big( G_{ab}^R - G_{ba}^{R\ast} \big) n_b^\ast 
&\!\!\!=\!\!\!& 
2i \big( {{\rm Im}}\, G_{xx}^R - {{\rm Re}}\, G_{xy}^R \big), 
\\
n_a^\ast \big( G_{ab}^R - G_{ba}^{R\ast} \big) n_b 
&\!\!\!=\!\!\!& 
2i \big( {{\rm Im}}\, G_{xx}^R + {{\rm Re}}\, G_{xy}^R \big). 
\ea
\end{subequations} 

\section{Enhancement of the resonant Casimir-Polder shift at $\widetilde{\omega}_{10}=1.9$} 
\label{sec:app-extra} 

\begin{figure}[h]
\centering
  \includegraphics[width=0.9\textwidth]{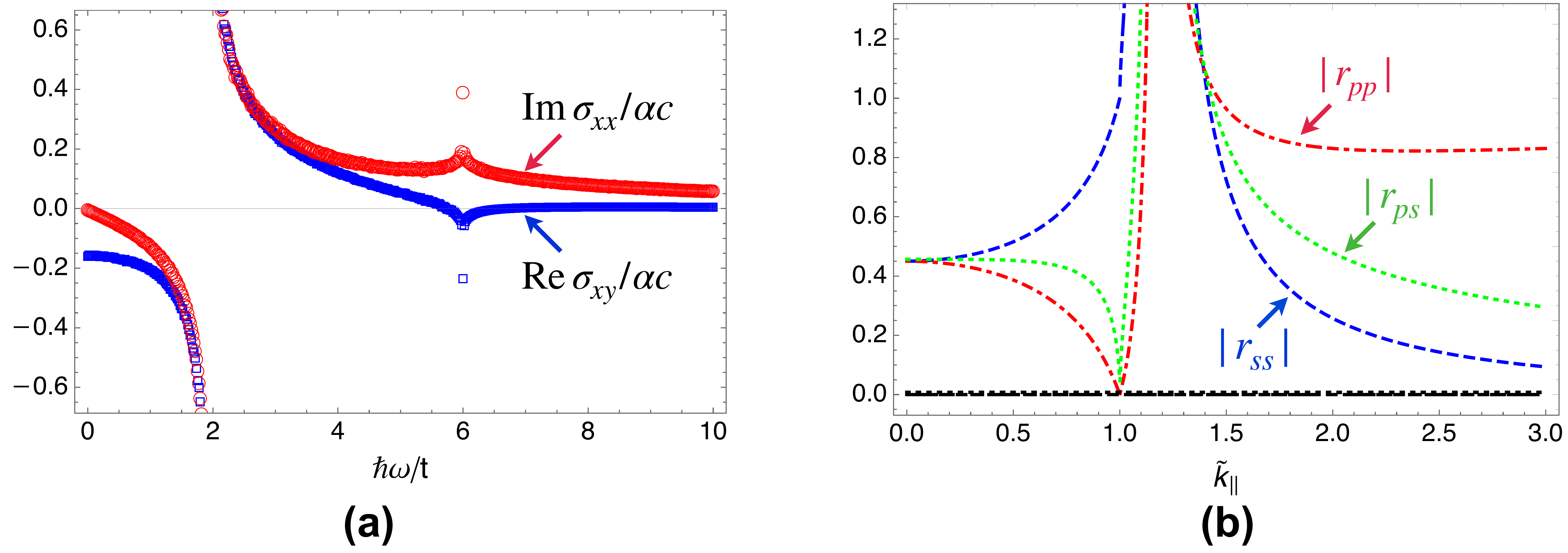}
  \caption{(a)~Frequency dependences of the imaginary part of $\widetilde{\sigma}_{xx}$ (red circles) and the real part of $\widetilde{\sigma}_{xy}$ (blue squares); 
  (b)~the $\widetilde{k}_\parallel$-dependence of $|r_{ss}|$ (blue, dashed), $|r_{pp}|$ (red, dot-dashed) and $|r_{ps}|=|r_{sp}|$ (green, dotted) for the dispersive case with $\widetilde{\omega}_{10} = 1.9$. The behaviors of the corresponding reflection coefficients in the nondispersive limit are plotted in black: $|r_{ss}|$ (dashed), $|r_{pp}|$ (dot-dashed), and $|r_{ps}|= |r_{sp}|$ (dotted).} 
  \label{fig:logdotplot}
\end{figure}
\begin{figure}[h]
\centering
  \includegraphics[width=0.48\textwidth]{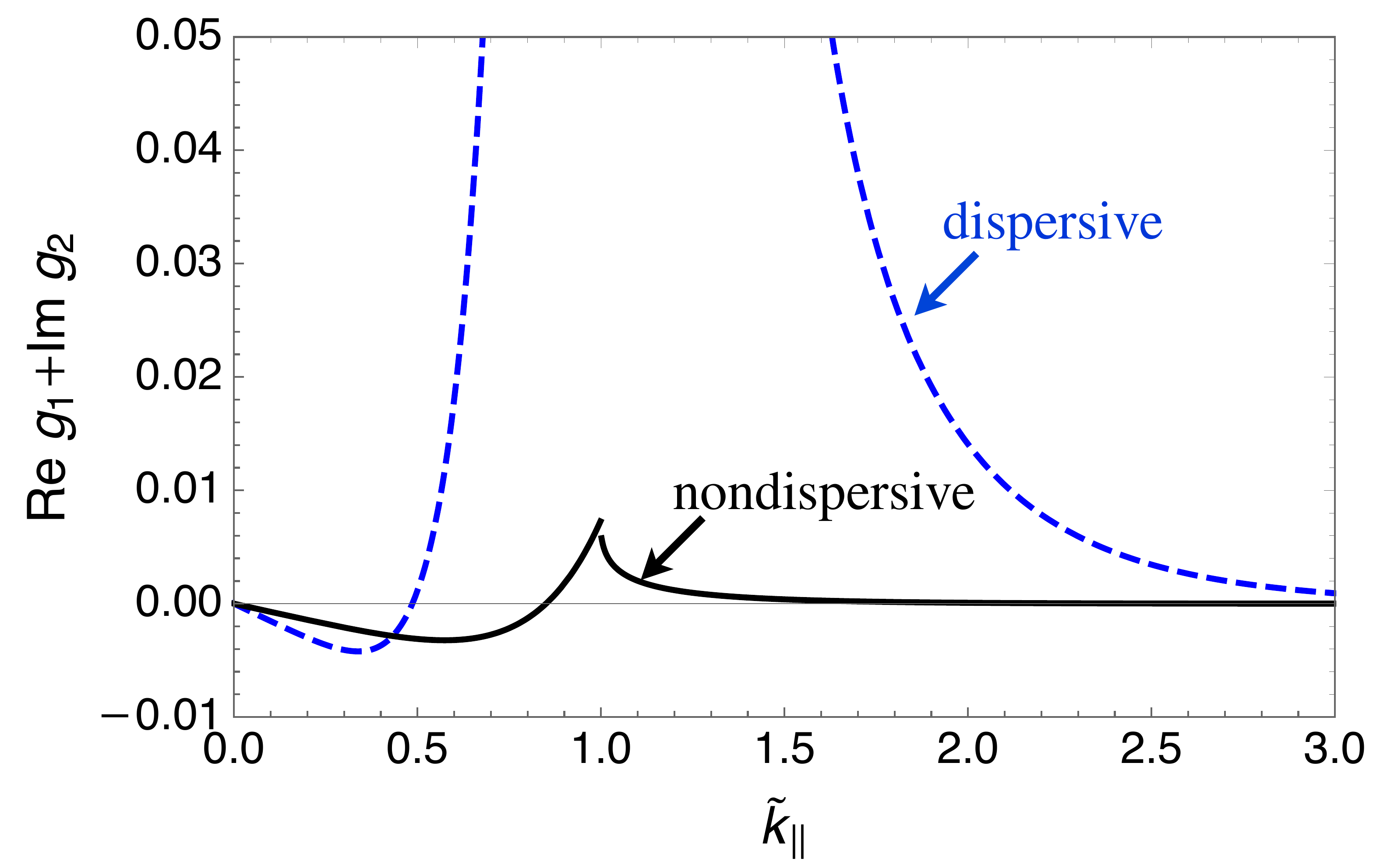}
  \caption{Behavior of ${{\rm Re}}\, g_1 + {{\rm Im}}\, g_2$ as a function of $\widetilde{k}_\parallel$ for the dispersive case with $\widetilde{\omega}_{10} = 1.9$ (blue) and the nondispersive limit (black). The functions $g_1$ and $g_2$ are defined in Eq.~(\ref{g1g2}).} 
  \label{fig:g}
\end{figure}

In this Appendix, we consider a circularly polarized dipole near a $C = -1$ Chern insulator, and look more deeply into how the resonant Casimir-Polder shift $\delta\widetilde{\omega}_{10}^{{\rm res}}$ becomes greatly enhanced at $\widetilde{\omega}_{10} = 1.9$, compared to the nondispersive limit.  
From Eq.~(\ref{formular}), the magnitude of $\delta\widetilde{\omega}_{10}^{{\rm res}}$ is determined by the magnitude of the reflection Green tensor, and from Eq.~(\ref{G-disp}), the latter's magnitude is in turn determined by the magnitude of the reflection coefficients. 

The magnitudes of the reflection coefficients depend on the frequency in the conductances $\widetilde{\sigma}_{xx}$ and $\widetilde{\sigma}_{xy}$. The frequency dependence of the conductances is studied and plotted in Sec.~II C of Ref.~\cite{lu2020}, where we found that for $\widetilde{\omega}_{10} < 2$, the real part of $\widetilde{\sigma}_{xx}$ and the imaginary part of $\widetilde{\sigma}_{xy}$ both vanish. For the reader's convenience, we quote here the formulaic results we found using the Kubo formula for the conductivity tensor~\cite{lu2020}: 
\begin{subequations}
\ba
\sigma_{xx}(\omega) 
&=& -\frac{1}{\hbar}
\int_{BZ} \!\frac{d^2\kv}{(2\pi)^2} 
{{\rm Re}} \, [\langle + | j_x | - \rangle \langle - | j_x | + \rangle]
\frac{f(d) - f(-d)}{2d}
\nonumber\\
&&\times
\bigg(
\pi \big(
\delta(\hbar\omega + 2d) + \delta(\hbar\omega - 2d)
\big)
+ 
\frac{2 i \hbar\omega}{\hbar^2\omega^2 - 4d^2} 
\bigg),
\\
\sigma_{xy}(\omega) 
&=& \frac{1}{\hbar}
\int_{BZ} \!\frac{d^2\kv}{(2\pi)^2} \, 
{{\rm Im}} \, [\langle + | j_x | - \rangle \langle - | j_y | + \rangle]
\frac{f(d) - f(-d)}{2d}
\nonumber\\
&&\times
\bigg(
\frac{4d}{4d^2-\hbar^2\omega^2} 
+
i \pi \big( 
\delta(\hbar\omega - 2d) - \delta(\hbar\omega + 2d)
\big)
\bigg), 
\ea
\label{cond-tensor}
\end{subequations}
where $d = \sqrt{d_x^2 + d_y^2 + d_z^2}$, $d_x = t \sin k_x a$, $d_y = t \sin k_y a$, $d_z = t (\cos k_x a + \cos k_y a) + u$, $f(d) = (\exp(\beta d) + 1)^{-1}$, and 
\begin{subequations}
\ba
&&{{\rm Im}}\, [\langle + | j_x | - \rangle \langle - | j_y | + \rangle] 
= - {{\rm Im}}\, [\langle - | j_x | + \rangle \langle + | j_y | - \rangle] 
\nonumber\\
&\!\!=\!\!& 
\frac{t^3 (ae)^2}{4\,d (d_x^2+d_y^2)} 
\Big(
\big( \cos k_x a \sin^2 k_y a + \cos k_y a \sin^2 k_x a \big) 
\big( 2(d_x^2+d_y^2 + t^2) - t^2 (\cos 2k_x a + \cos 2k_y a) \big)
\nonumber\\
&&\qquad\qquad+ 
4td_z \cos k_x a \cos k_y a \big( \sin^2 k_x a + \sin^2 k_y a \big)
\Big), 
\\
&&{{\rm Re}}\, [\langle + | j_x | - \rangle \langle - | j_x | + \rangle] 
\nonumber\\
&\!\!=\!\!& 
\frac{(tae)^2}{16\,d^2 (d_x^2+d_y^2)} 
\Big(
\sin^2 k_x a
\big(
t^2 (\cos 2k_x a + \cos 2k_y a) - 2((d_x^2+d_y^2) + t^2) 
\big)
\nonumber\\
&&\quad\times
\big(
t^2 (\cos 2k_x a + \cos 2k_y a) - 2(d_x^2+d_y^2 + t^2) - 8 t d_z \cos k_x a
\big)
\nonumber\\
&&\qquad\qquad+ 4 t^2 d_z^2 \sin^2 2k_x a
+ 16 t^2 d^2 \cos^2 k_x a \sin^2 k_y a
\Big). 
\ea
\end{subequations}
As in Ref.~\cite{lu2020}, we show in Fig.~\ref{fig:logdotplot}(a) the plots of the frequency dependence of the conductivity components which are non-vanishing, i.e., the imaginary part of $\widetilde{\sigma}_{xx}$ (red circles) and the real part of $\widetilde{\sigma}_{xy}$ (blue squares). As we explained in Ref.~\cite{lu2020}, for $\widetilde{\omega}_{10} < 2$, ${{\rm Im}} \, \widetilde{\sigma}_{xx}$ is zero and ${{\rm Re}} \, \widetilde{\sigma}_{xy} = -\alpha \approx -1/137$, a feature that we also see in the plot. 

Figure~\ref{fig:logdotplot}(b) shows plots for the $\widetilde{k}_\parallel$-dependence of the magnitudes of the reflection coefficients for the dispersive case where $\widetilde{\omega}=1.9$, and also plots for the magnitudes of the reflection coefficients corresponding to the nondispersive limit. 
In the nondispersive limit, the reflection coefficients have constant values, with $r_{ss} = - r_{pp} \approx -\alpha^2 \sim \mathcal{O}(10^{-4})$ and $r_{ps} = r_{sp} \approx \alpha \sim \mathcal{O}(10^{-2})$. On the other hand, we see that the reflection coefficients for the dispersive case with $\widetilde{\omega}_{10} = 1.9$ depend on $\widetilde{k}_\parallel$, and are generally much larger than those in the nondispersive limit. 

Finally, let us express $\widetilde{G}_{xx}^R$ and $\widetilde{G}_{xy}^R$ in the form
\be
\widetilde{G}_{xx}^R = \int_0^\infty \!\!\! d\widetilde{k}_\parallel \, g_1(\widetilde{k}_\parallel),
\quad
\widetilde{G}_{xy}^R = \int_0^\infty \!\!\! d\widetilde{k}_\parallel \, g_2(\widetilde{k}_\parallel),
\label{g1g2}
\ee
which allows us to express Eq.~(\ref{E1cl-circ}) in the form
\be
\delta\widetilde{\omega}_{10}^{{\rm res}} 
=
- \frac{3}{4} \int_0^\infty \!\!\! d\widetilde{k}_\parallel 
\Big( 
{{\rm Re}}\, g_1(\widetilde{k}_\parallel) + {{\rm Im}}\, g_2(\widetilde{k}_\parallel) 
\Big). 
\ee
In Fig.~\ref{fig:g}, we plot the behavior of ${{\rm Re}}\, g_1(\widetilde{k}_\parallel) + {{\rm Im}}\, g_2(\widetilde{k}_\parallel)$ for positive values of $\widetilde{k}_\parallel$, for both the dispersive case with $\widetilde{\omega}_{10} = 1.9$ (blue) and the nondispersive limit (black). We see that over the range plotted, ${{\rm Re}}\, g_1(\widetilde{k}_\parallel) + {{\rm Im}}\, g_2(\widetilde{k}_\parallel)$ for the dispersive case generally has a much larger magnitude than the nondispersive limit. For values of $\widetilde{k}_\parallel$ much larger than 1, ${{\rm Re}}\, g_1(\widetilde{k}_\parallel) + {{\rm Im}}\, g_2(\widetilde{k}_\parallel)$ becomes rapidly exponentially suppressed. 

To summarize: in the nondispersive limit, $\widetilde{\sigma}_{xx} = 0$ and $\widetilde{\sigma}_{xy} \approx -1/137$, whereas the magnitudes of both $\widetilde{\sigma}_{xx}$ and $\widetilde{\sigma}_{xy}$ are of the order of unity in the dispersive case with $\widetilde{\omega}_{10} = 1.9$. This dispersion contributes to an enhancement in the reflection coefficients, which directly results in the enhancement of $\delta\widetilde{\omega}_{10}^{{\rm res}}$. 

\section{Asymptotic regimes for the resonant Casimir-Polder shift} 
\label{sec:app3} 

In this Appendix, we obtain the asymptotic behavior of the resonant Casimir-Polder shift $\delta\widetilde{\omega}_{10}^{{\rm res}}$ in the near-field ($\eta \to 0$) and far-field limit $\eta \gg 1$. For the dipole transitions which are polarized perpendicular to the surface, parallel to the surface, and right circularly polarized in the plane of the surface, this amounts to looking at the asymptotic behavior of ${{\rm Re}}\,G_{xx}^R$, ${{\rm Re}}\,G_{zz}^R$ and ${{\rm Im}}\,G_{xy}^R$. 
For our asymptotic analysis, we follow the same procedure as the one described in Ref.~\cite{lu2020}. Firstly, we scale out the dependence on $\eta$ in the exponential factors in Eqs.~(\ref{dyadic}) by defining a new dimensionless variable $t \equiv \tk_z \eta$ for the range $0 \leq \tk_\parallel < 1$ (recalling that $\tk_z = (1 - \tk_\parallel^2)^{1/2}$), whence $\tk_\parallel d\tk_\parallel = - \tk_z d\tk_z = - t \, dt/\eta^2$. For this range, $0 \leq \tk_z < 1$ and thus $0 \leq t < \eta$. 
Then for the range $1 \leq \tk_\parallel < \infty$, we have $0 \leq \tk_z < i\infty$. If we define $\tk_z \equiv i \ell$, then $0 \leq \ell < \infty$, and we can define $t \equiv \ell \eta$, whereupon $\tk_\parallel d\tk_\parallel = \ell \, d\ell = t \, dt/\eta^2$. 
After rescaling and using Eqs.~(\ref{r-coeffs}), Eqs.~(\ref{dyadic}) become
\begin{subequations}
\ba
\tG_{xx}^R(\rv_0,\rv_0;\omega_{10})
&\!\!=\!\!& 
g_{1}(\rv_0,\rv_0; \omega_{10}) + h_{1}(\rv_0,\rv_0; \omega_{10}), 
\nonumber\\
g_{1}(\rv_0,\rv_0; \omega_{10}) &\!\!=\!\!& 
- \frac{i}{2\eta} 
\int_0^\eta \!\!dt 
\frac{
  [(1 + (\frac{t}{\eta})^2)(\widetilde{\sigma}_{xx}^2 + \widetilde{\sigma}_{xy}^2) + (\frac{\eta}{t} + (\frac{t}{\eta})^3)\widetilde{\sigma}_{xx}]e^{it}
  }
  {1 + \widetilde{\sigma}_{xx}^2 + \widetilde{\sigma}_{xy}^2 + (\frac{\eta}{t} + \frac{t}{\eta})\widetilde{\sigma}_{xx}}, 
\nonumber\\
h_{1}(\rv_0,\rv_0; \omega_{10})
&\!\!=\!\!& 
-
\frac{1}{2\eta} 
\int_0^\infty \!\!\!dt 
\frac{[(1 - (\frac{t}{\eta})^2)(\widetilde{\sigma}_{xx}^2 + \widetilde{\sigma}_{xy}^2) - i ((\frac{t}{\eta})^3 + \frac{\eta}{t})\widetilde{\sigma}_{xx}]e^{-t}}
{1 + \widetilde{\sigma}_{xx}^2 + \widetilde{\sigma}_{xy}^2 + i (\frac{t}{\eta} - \frac{\eta}{t})\widetilde{\sigma}_{xx}};
\\ 
\nonumber \\ 
\tG_{xy}^R(\rv_0,\rv_0;\omega_{10}) 
&\!\!=\!\!&   
\tG_{yx}^R(\rv_0,\rv_0;\omega_{10}) 
=
g_{2}(\rv_0,\rv_0; \omega_{10}) + h_{2}(\rv_0,\rv_0; \omega_{10}), 
\nonumber\\
g_{2}(\rv_0,\rv_0; \omega_{10})
&\!\!=\!\!& 
- \frac{i}{\eta^2} 
\int_0^\eta \!\!dt \, t 
\frac{\widetilde{\sigma}_{xy}}{1 + \widetilde{\sigma}_{xx}^2 + \widetilde{\sigma}_{xy}^2 + (\frac{\eta}{t} + \frac{t}{\eta})\widetilde{\sigma}_{xx}} e^{it}, 
\nonumber\\
h_{2}(\rv_0,\rv_0; \omega_{10})
&\!\!=\!\!& 
- \frac{i}{\eta^2} \int_0^\infty \!\!dt \, t \frac{\widetilde{\sigma}_{xy}}{1 + \widetilde{\sigma}_{xx}^2 + \widetilde{\sigma}_{xy}^2 + i (\frac{t}{\eta} - \frac{\eta}{t})\widetilde{\sigma}_{xx}} e^{-t};
\\ 
\nonumber \\ 
\tG_{zz}^R(\rv_0,\rv_0;\omega_{10}) 
&\!\!=\!\!& 
g_{3}(\rv_0,\rv_0; \omega_{10}) + h_{3}(\rv_0,\rv_0; \omega_{10}), 
\nonumber\\
g_{3}(\rv_0,\rv_0; \omega_{10})
&\!\!=\!\!& 
\frac{i}{\eta} \int_0^\eta \!\!dt \, 
\bigg( 1 - \frac{t^2}{\eta^2} \bigg)
 \frac{\widetilde{\sigma}_{xx}^2 + \widetilde{\sigma}_{xy}^2 + \frac{t}{\eta} \widetilde{\sigma}_{xx}}
 {1 + \widetilde{\sigma}_{xx}^2 + \widetilde{\sigma}_{xy}^2 + (\frac{\eta}{t} + \frac{t}{\eta})\widetilde{\sigma}_{xx}} 
 e^{it}, 
\nonumber\\
h_{3}(\rv_0,\rv_0; \omega_{10})
&\!\!=\!\!& 
\frac{1}{\eta} \int_0^\infty \!\!dt \, 
\bigg( 1 + \frac{t^2}{\eta^2} \bigg)
 \frac{\widetilde{\sigma}_{xx}^2 + \widetilde{\sigma}_{xy}^2 + i \frac{t}{\eta} \widetilde{\sigma}_{xx}}
 {1 + \widetilde{\sigma}_{xx}^2 + \widetilde{\sigma}_{xy}^2 + i (\frac{t}{\eta} - \frac{\eta}{t})\widetilde{\sigma}_{xx}} 
 e^{-t}. 
\ea
\label{C1}
\end{subequations}
In the above, we have expressed each Green tensor component as the sum of a contribution with an oscillatory integrand, $g(\rv_0,\rv_0; \omega_{10})$, and a contribution with an exponentially decaying integrand, $h(\rv_0,\rv_0; \omega_{10})$.

\subsection{Near-field asymptotics}
\label{app:nearfield}

Let us consider the asymptotic behavior of the resonant Casimir-Polder shift in the near-field limit in the low and high frequency regimes. This amounts to considering the behavior of the
contributions ${{\rm Re}}\,\tG_{xx}^R = {{\rm Re}}\, g_1 + {{\rm Re}}\, h_1$, ${{\rm Im}}\,\tG_{xy}^R = {{\rm Im}}\, g_2 + {{\rm Im}}\, h_2$ and ${{\rm Re}}\,\tG_{zz}^R = {{\rm Re}}\, g_3 + {{\rm Re}}\, h_3$ to the terms in Eq.~(\ref{C1}) in the limit that $\eta \to 0$. 

As the resonant Casimir-Polder shift for our considered dipole polarizations involves $\tG_{xx}^R$, $\tG_{xy}^R$ and $\tG_{zz}^R$, let us derive the near-field limits for the functions $g_1$, $g_2$, $g_3$, $h_1$, $h_2$ and $h_3$, for the case that the de-excitation frequency is not close to frequencies associated with van Hove singularities. 
Since $\eta$ is the upper limit in the integration over $t$ in the functions $g_1$, $g_2$ and $g_3$ in Eqs.~(\ref{C1}), we have that $t/\eta \to 1$, $\eta/t \to 1$ and $e^{it} \to 1$ as $\eta \to 0$. We obtain 
\begin{subequations}
\ba
g_1(\rv_0,\rv_0;\omega_{10}) 
&\!\!\approx\!\!& 
  -\frac{i}{2\eta} \int_0^\eta \!\!dt\, 
  \frac{2(\widetilde{\sigma}_{xx}^2 + \widetilde{\sigma}_{xy}^2) + 2\widetilde{\sigma}_{xx}}
  {1 + \widetilde{\sigma}_{xx}^2 + \widetilde{\sigma}_{xy}^2 + 2\widetilde{\sigma}_{xx}} 
  =
  - \frac{i(\widetilde{\sigma}_{xx}^2 + \widetilde{\sigma}_{xy}^2 + \widetilde{\sigma}_{xx})}
  {1 + \widetilde{\sigma}_{xx}^2 + \widetilde{\sigma}_{xy}^2 + 2\widetilde{\sigma}_{xx}}, 
\\
g_2(\rv_0,\rv_0;\omega_{10}) 
&\!\!\approx\!\!&
- \int_0^\eta \!\!dt
\frac{i}{\eta^2} \frac{t \widetilde{\sigma}_{xy}}{1 + \widetilde{\sigma}_{xx}^2 + \widetilde{\sigma}_{xy}^2 + 2\widetilde{\sigma}_{xx}} 
=
-\frac{i}{2} \frac{\widetilde{\sigma}_{xy}}{1 + \widetilde{\sigma}_{xx}^2 + \widetilde{\sigma}_{xy}^2 + 2\widetilde{\sigma}_{xx}}, 
\\
g_3(\rv_0,\rv_0;\omega_{10}) 
&\!\!\approx\!\!&
  \frac{i}{\eta} 
  \int_0^\eta \!\!dt \, 
  \frac{\widetilde{\sigma}_{xx}^2 + \widetilde{\sigma}_{xy}^2 + \widetilde{\sigma}_{xx}}{1 + \widetilde{\sigma}_{xx}^2 + \widetilde{\sigma}_{xy}^2 + 2\widetilde{\sigma}_{xx}} 
  \Big(1 - \frac{t^2}{\eta^2}\Big)  
  =
  \frac{2i}{3} 
  \frac{\widetilde{\sigma}_{xx}^2 + \widetilde{\sigma}_{xy}^2 + \widetilde{\sigma}_{xx}}{1 + \widetilde{\sigma}_{xx}^2 + \widetilde{\sigma}_{xy}^2 + 2\widetilde{\sigma}_{xx}}. 
  \ea
\end{subequations}
Next, let us consider the near-field limits for the functions $h_1$, $h_2$ and $h_3$. We can perform a perturbation expansion in powers of $\eta/t$ in the integrand and retain the leading order term. This leads to
\begin{subequations}
\ba
h_1(\rv_0,\rv_0;\omega_{10}) 
&\!\!\approx\!\!& 
\frac{1}{2\eta^3} 
\int_0^\infty \!\!\! dt \, 
t^2 e^{-t} 
= \frac{1}{\eta^3}, 
\\
h_2(\rv_0,\rv_0;\omega_{10}) 
&\!\!\approx\!\!& 
- \frac{1}{\eta} 
\int_0^\infty \!\!\! dt \, 
\frac{\tsigma_{xy}}{\tsigma_{xx}} e^{-t} 
= 
- 
\left( 
\frac{\tsigma_{xy}}{\tsigma_{xx}} 
\right)
\frac{1}{\eta}, 
\\
h_3(\rv_0,\rv_0;\omega_{10}) 
&\!\!\approx\!\!& 
\frac{1}{\eta^3} \int_0^\infty dt\, t^2 e^{-t} = \frac{2}{\eta^3}, 
\ea
\end{subequations}
From the above, we see again that the near-field limiting values of ${{\rm Re}}\,h_1$, ${{\rm Im}}\,h_2$ and ${{\rm Re}}\,h_3$ dominate the limiting values of ${{\rm Re}}\,g_1$, ${{\rm Im}}\,g_2$ and ${{\rm Re}}\,g_3$, which implies that 
${{\rm Re}}\,\tG_{xx}^R \approx {{\rm Re}}\, h_1$, ${{\rm Im}}\,\tG_{xy}^R \approx {{\rm Im}}\, h_2$ and ${{\rm Re}}\,\tG_{zz}^R \approx {{\rm Re}}\, h_3$. 

Using Eq.~(\ref{E1cl-perp}), we obtain the near-field limit of the resonant Casimir-Polder shift for a dipole transition polarized perpendicular to the surface:
\be
\delta\widetilde{\omega}_{10}^{{\rm res}} 
\approx 
-\frac{3}{4} 
{{\rm Re}} \, 
h_3(\rv_0,\rv_0;\omega_{10}) 
\approx 
-\frac{3}{2\eta^3} \,, 
\ee
Similarly, using Eq.~(\ref{E1cl-para}) we obtain the near-field limit of the resonant Casimir-Polder shift for a dipole transition polarized parallel to the surface:
\be
\delta\widetilde{\omega}_{10}^{{\rm res}} 
\approx 
-\frac{3}{4} 
{{\rm Re}} \, 
h_1(\rv_0,\rv_0;\omega_{10}) 
\approx 
-\frac{3}{4\eta^3} \,.
\ee
Finally, the near-field limit of the resonant Casimir-Polder shift for a right circular dipole polarization, Eq.~(\ref{E1cl-circ}), is 
\ba 
\delta\widetilde{\omega}_{10}^{{\rm res}} 
\approx 
-\frac{3}{4} 
\left(
{{\rm Re}} \, 
h_1(\rv_0,\rv_0;\omega_{10}) 
+
{{\rm Im}} \, 
h_2(\rv_0,\rv_0;\omega_{10}) 
\right)
\approx -\frac{3}{4\eta^3}.
\ea
The above near-field limits are the same for all the frequency regimes. 

\subsection{Far-field asymptotics}
\label{app:farfield}

In the far-field limit, $t/\eta \ll 1$. We can perform a perturbation expansion in powers of $t/\eta$ in the integrands of the functions $g_1$, $g_2$, $g_3$, $h_1$, $h_2$ and $h_3$ from Eqs.~(\ref{C1}), and retain the leading order term. We obtain 
\begin{subequations}
\label{gh-farfield}
\ba
g_1(\rv_0,\rv_0;\omega_{10}) 
&\!\!\approx\!\!& 
-\frac{i}{2\eta} \int_0^\eta \!\!\!dt \, e^{it} = \frac{1}{2\eta} \left( 1 - \cos\eta - i \sin\eta \right),
\\
g_2(\rv_0,\rv_0;\omega_{10}) 
&\!\!\approx\!\!&
-\frac{i}{\eta^3} \int_0^\eta \!\!\!dt \, t^2 e^{it} \frac{\tsigma_{xy}}{\tsigma_{xx}} 
= - \frac{1}{\eta^3} \frac{\tsigma_{xy}}{\tsigma_{xx}}  
\left( 2 + \left( \eta^2 -2 + 2 i \eta \right) e^{i \eta} \right),
\\
g_3(\rv_0,\rv_0;\omega_{10}) 
&\!\!\approx\!\!&
\frac{i}{\eta^2} \int_0^\eta \!\!\!dt \, t \, e^{it} 
\left( \frac{\tsigma_{xx}^2 + \tsigma_{xy}^2}{\tsigma_{xx}} \right)
= 
- \frac{i}{\eta^2} 
\left( \frac{\tsigma_{xx}^2 + \tsigma_{xy}^2}{\tsigma_{xx}} \right)
\left( 1 - \left( 1 - i \eta \right) e^{i \eta} \right), 
\\
h_1(\rv_0,\rv_0;\omega_{10}) 
&\!\!\approx\!\!& 
-\frac{1}{2\eta} \int_0^\infty \!\!\!dt \, e^{-t} 
= 
- \frac{1}{2\eta}, 
\\
h_2(\rv_0,\rv_0;\omega_{10}) 
&\!\!\approx\!\!& 
\frac{1}{\eta^3} \int_0^\infty \!\!\!dt \, t^2 e^{-t} \left( \frac{\tsigma_{xy}}{\tsigma_{xx}} \right)
= 
\frac{2}{\eta^3} \left( \frac{\tsigma_{xy}}{\tsigma_{xx}} \right), 
\\
h_3(\rv_0,\rv_0;\omega_{10}) 
&\!\!\approx\!\!& 
\frac{i}{\eta^2} \int_0^\infty \!\!\!dt \, t \, e^{-t} 
\left( \frac{\tsigma_{xx}^2 + \tsigma_{xy}^2}{\tsigma_{xx}} \right)
=
\frac{i}{\eta^2} 
\left( \frac{\tsigma_{xx}^2 + \tsigma_{xy}^2}{\tsigma_{xx}} \right).   
\ea
\end{subequations}
As $\tG^R = g + h$, we have 
\begin{subequations}
\ba
\tG_{xx}^R(\rv_0,\rv_0;\omega_{10}) 
&\!\!\approx\!\!& 
- \frac{\cos\eta + i \sin\eta}{2\eta} 
,
\\
\tG_{xy}^R(\rv_0,\rv_0;\omega_{10}) 
&\!\!\approx\!\!& 
- \frac{1}{\eta^3} \frac{\tsigma_{xy}}{\tsigma_{xx}}  
 \left( \eta^2 -2 + 2 i \eta \right) e^{i \eta} ,
\\
\tG_{zz}^R(\rv_0,\rv_0;\omega_{10}) 
&\!\!\approx\!\!& 
\frac{i}{\eta^2} 
\left( \frac{\tsigma_{xx}^2 + \tsigma_{xy}^2}{\tsigma_{xx}} \right)
\left( 1 - i \eta \right) e^{i \eta}.  
\ea
\label{G-farfield}
\end{subequations}
In the low frequency regime ($\omega_{10} < 2(2t-|u|)/\hbar $) and high frequency regime ($\omega_{10} > 2(2t-|u|)/\hbar $), $\widetilde{\sigma}_{xx} = i\widetilde{\sigma}_{xx}''$ and $\widetilde{\sigma}_{xy} = \widetilde{\sigma}_{xy}'$~\cite{lu2020}. 
The resonant Casimir-Polder shifts for our considered dipole polarizations involve ${{\rm Re}}\, \tG_{xx}^R$, ${{\rm Im}}\, \tG_{xy}^R$ and ${{\rm Re}}\, \tG_{zz}^R$. 
Using~(\ref{G-farfield}), we obtain their far-field limits: 
\begin{subequations}
\ba
{{\rm Re}}\, \tG_{xx}^R(\rv_0,\rv_0;\omega_{10}) 
&\!\!\approx\!\!& 
- \frac{\cos\eta}{2\eta} 
,
\\
{{\rm Im}}\, \tG_{xy}^R(\rv_0,\rv_0;\omega_{10}) 
&\!\!\approx\!\!& 
- \frac{1}{\eta^3} \frac{\tsigma_{xy}'}{\tsigma_{xx}''}  
\left( 2\cos\eta + 2\eta \sin \eta - \eta^2\cos\eta \right),
\\
{{\rm Re}}\, \tG_{zz}^R(\rv_0,\rv_0;\omega_{10}) 
&\!\!\approx\!\!& 
- \frac{1}{\eta^2} 
\left( \frac{(\tsigma_{xx}'')^2 - (\tsigma_{xy}')^2}{\tsigma_{xx}''} \right)
\left( 
\cos\eta + \eta \sin\eta
\right).
\ea
\end{subequations}

\end{widetext}

\end{document}